\documentclass{ws-ijmpe}
\usepackage[super,compress]{cite}
\def\be{\begin{equation}}
\def\ee{\end{equation}}
\def\bea{\begin{eqnarray}} 
\def\eea{\end{eqnarray}}
\def\l{\label}

\def\hahat{\hat{H}}

\def\hahat0{\hat{H}_0}

\def\sin{\hbox{sin}}

\def\exp{\hbox{exp}}

\def\d{\hbox{d}}
\def\eps{\varepsilon}
\def\epsi{\mathcal{E}}

\def\siml{\hbox{\kern.1em \lower.6ex \hbox{$\sim$} \kern-1.12em
 \raise.6ex \hbox{$<$} \kern.1em}}
\def\simg{\hbox{\kern.1em \lower.6ex \hbox{$\sim$} \kern-1.12em
 \raise.6ex \hbox{$>$} \kern.1em}}
\begin{document}

\markboth{A.G.Magner, S.P.Maydanyuk, A.Bonasera, H.Zheng, A.I.Levon, T.Depastas, U.V.Grygoriev}{Neutron stars as dense liquid drop at equilibrium}

\catchline{}{}{}{}{}

\title{NEUTRON STARS AS DENSE LIQUID DROP AT EQUILIBRIUM
  WITHIN THE EFFECTIVE SURFACE
 APPROXIMATION}

  \author{A.G.~Magner}
  \address{{Nuclear Theory Department,
      Institute for Nuclear Research, 
      03028 Kyiv, Ukraine\\
    Cyclotron Institute, Texas A\&M University,
  College Station, Texas 77843, USA}\\
        first\_alexander.magner66@gmail.com}

  \author{S.P.~Maydanyuk}
  \address{{Nuclear Processes Department,
      Institute for Nuclear Research,  
      03028 Kyiv, Ukraine}\\
   Institute of Modern Physics, Chinese Academy of Sciences, Lanzhou,
730000, China\\
second\_sergei.maydanyuk@wigner.hu}

  \author{A.~Bonasera}
\address{{Cyclotron Institute, Texas A\&M University,
  College Station, Texas 77843, USA}\\  
third\_abonasera@me.com}

   \author{H.~Zheng}
   \address{{School of Physics and Information Technology,
       Shaanxi Normal University,Xi'an 710119, China}\\  
fourth\_zhengh@snnu.edu.cn}

      \author{A.I.~Levon}
      \address{{Nuclear Reactions Department,
          Institute for Nuclear Research, 
          03028 Kyiv, Ukraine}\\  
fifth\_alevon78@gmail.com}

  \author{T.M.~Depastas}
\address{{Cyclotron Institute, Texas A\&M University,
  College Station, Texas 77843, USA}\\  
sixth\_tdepastas@tamu.edu}

  \author{U.V.~Grygoriev}
  \address{{Nuclear Theory Department,
      Institute for Nuclear Research, 
      03028 Kyiv, Ukraine}\\  
seventh\_Yustim@ukr.net}  

\maketitle

\begin{history}
\received{Day Month Year}
\revised{Day Month Year}
\end{history}

\begin{abstract}
 The macroscopic model is formulated for a neutron star (NS) as a
 perfect liquid drop at the equilibrium. We use 
 the leptodermic approximation $a/R\ll 1$, where
 $a$ is the crust thickness of the effective NS surface (ES), and $R$ is the mean radius of the ES curvature. 
 Within the approximate Schwarzschild metric solution to the general relativity theory equations for the spherically symmetric systems,
 the macroscopic gravitation is taken into account in terms of the total separation particle energy and incompressibility.
  Density distribution $\rho$ across the ES in the normal direction to the ES was obtained analytically for a general form of the energy density
  $\mathcal{E}(\rho)$.
  For the typical crust thickness, and effective radius, one finds  the leading expression for the density $\rho$. NS masses are analytically calculated as a sum of the volume
  and surface terms, taking into account the radial curvature of the
  metric space, in reasonable agreement with the recently measured
 masses for several neutron stars.  We
 derive the simple macroscopic equation of state (EoS)
with the surface correction. The analytical and numerical solutions to Tolman-Oppenheimer-Volkoff equations for the pressure are in good agreement with the volume part of our EoS. 
\end{abstract}

\keywords{Neutron stars; dense liquid drop;
  Schwarzschild metric; energy density; effective surface;
  equation of state.}

\ccode{PACS number: 21.65.Mn,26.60.Gj}


\section{INTRODUCTION}
\label{introd}

R.C. Tolman suggested \cite{RT39} first to study the simplest model
for a neutron star (NS)
considering it as a dense liquid-matter drop at its
equilibrium under
the gravitational, nuclear and other realistic
  forces;
see also his book \cite{RT87}, chapt. 7, sect. 96.
Within this model, the Einstein-Gilbert equations of the
    General Relativistic Theory (GRT)
    for the spherical
symmetry case has been reduced to the three independent equations for
four unknown
quantities, 
namely, two parameters,
    $\lambda$ and $\nu$, of the Schwarzschild metric
    (see
also Ref.~\citen{LLv2}, the chapters 11 and 12)
\be\l{Schwarzgen}
    {\rm d} s^2=-e^\lambda{\rm d} r^2-
     r^2 {\rm d}\theta^2 -
     r^2 \sin^2\theta {\rm d} \phi^2
     + e^\nu c^2{\rm d} t^2,
     \ee
    and 
    the pressure $P$, and energy density
    $\epsi$.
    To get the complete system of equations Tolman suggested also
    to derive
independently
the equation of
state (EoS), $\epsi=\epsi(\rho)$, or $P=P(\rho)$.
The EoS can be found
from the condition of a static equilibrium for a 
liquid-matter drop under gravitational, and nuclear (and Coulomb)
forces, which all should be taken into account by the
simplified GRT equations for the gravitational field.
Following Tolman's ideas, Oppenheimer and Volkoff\cite{OV39} 
derived the simple (TOV)
equations by using essentially
the macroscopic properties of the system
[Refs.~\citen{LLv2} (chapt. 12) and \citen{LLv6} (chapt. 1)]
    of the
    system as a perfect liquid drop at equilibrium.
As shown in 
Ref.~\citen{RT87}, one can derive the analytical solution
for the pressure $P$ as function
of the radial coordinate.  So far, the 
TOV equations \cite{OV39} are considered with  
the equation of 
    state, $\epsi=\epsi(\rho)$, which was obtained independently
of the macroscopic assumptions used in the TOV derivations.
For instance, the EoS with a polytropic expression for the
pressure\cite{ST04,BK11}
    $P=P(\rho)$ 
    as function of the density $\rho$ is assumed to be similar to that for
   a particle gas system with fitted parameters.
    The gradient terms are usually neglected
    in the energy density, and then, no equilibrium for
    a finite gas system with no external fields. 
However, the pressure $P$ was 
calculated as
function of the radial $r$ coordinate
    by using the TOV equations for another dense
    liquid-drop system. Such a system, including the gravitation,
    can be described  mainly at a stable equilibrium
    by short-range forces, in contrast to a gas matter
    with long-range inter-particle interaction. Then,
the NS mass $M$ was obtained numerically as function of the
NS radius $R$, $M=M(R)$,  with 
important restrictions 
due to those of the NS radius. 
    In any case, we should emphasize
    the
    importance of the gradient terms in the energy density $\epsi(\rho)$
    for a macroscopic condition of the system equilibrium, in spite of a
   relative small NS crust thickness.    
As a hint to these conclusions, see many applications
 of the
TOV equations, e.g., in
Refs.~\citen{WF88,CB97,AA98,SH06,Ko08,CH08,SG13,BZ14,BC15,LH19,SBL23}.
Concerning the relation of the nuclear and neutron liquid-drop
    models (LDMs)
    to NS properties,
we should also mention the work\cite{BBP71} by
Baym, Bethe, and Pethick. They discuss the
liquid matter drop with the leptodermic property
having a sharp decrease of the particle
number density
in a relatively small edge region considered as its surface.
    Clear specific definitions and complete updated results
    for the energy density with
density gradient (surface) terms and  for equations of
the infinite matter state
 with many inter-particle
forces in the non-relativistic and relativistic cases
can be found in the 
recent review Ref.~\citen{SBL23}.

\vspace{0.2cm}
 \begin{figure*}[!ht]
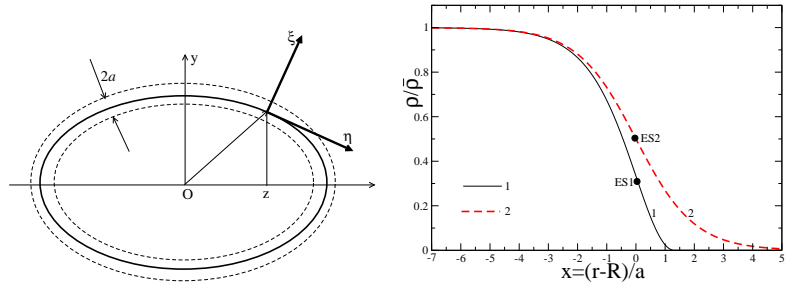

   \vskip1mm
  \centerline{\includegraphics[width=4.9cm]{fig1a_jmpe.eps}
   ~
  \includegraphics[width=5.1cm]{Fig1b-den_a1R10.eps}}

   \vspace{-0.5cm}
   \vskip4mm\caption{{\small
        {\it Left:} A qualitative plot 
       for the deformed ES with the local
       coordinate system $\xi,\eta$, where
 $\xi$ is the axis perpendicular, and $\eta$ is that parallel to the ES. 
       We show also a full diffuse surface thickness $2a$.
       {\it Right:}
         Particle number density, $\rho$, in units of the saturation value,
       $\overline{\rho}$,
       as function of the variable $x=\xi/a=(r-R)/a$ 
       for the 
       NS 
      in a simple compressed
       LDM at the stable
       equilibrium.
       Solid line is related to the asymmetric solution,
       Eq.~(\ref{b0g0}),
       for $\beta=\gamma=0$ (line ``1''). 
       Dashed line presents the same but for the Wilets symmetric solution,
       Eq.~(\ref{wilets}) (line ``2'').
       Parameters, for example:
          the effective radius $R=10$ km, diffuseness of the NS crust
           $a=1$ km. 
         The dots ES1 and ES2 show the ES at the effective radius $R$ for
         the solid,
      Eq.~(\ref{b0g0}), 
      and dashed, Eq.~(\ref{wilets}), lines, respectively.
      }}
\label{fig1}
\end{figure*}
 Taking Tolman's ideas, one can try first to extend the
 EoS
 to those for a dense macroscopic\cite{LLv2,LLv6}
 system of particles. 
 In this leptodermic
 system, the particle density $\rho$ is function of the
 radial coordinate with
exponentially decreasing behavior 
from an almost constant saturation value inside of the dense system
to that through the effective
surface (ES) of NS in a relatively small crust range $a$;
see Fig.~\ref{fig1}.
The ES is defined 
as the points of spatial coordinates with 
    a maximum density gradient.
To obtain the analytical solutions for the
particle density and EoS, we 
use the
 leptodermic
    approximation\cite{LLv2,RT87,OV39,SBL23,LP01,HP01PRL,HP01PRC,PH00Poland,ABCG09,FG23,DFG23,BBP71,HPY07,LH19,ST04}, $a/R \ll 1$,
where  $R$ is the (curvature) radius of the
ES.

 Within this effective surface 
approximation (ESA), $a/R \ll 1$,
simple and
accurate solutions of many nuclear and liquid-drop
problems involving the particle number
density distributions 
were obtained 
for nuclei\cite{wilets,strtyap,tyapin,strmagbr,strmagden,MS09,BM13,BM15}.
The ESA exploits
the property of saturation of the nuclear particle density $\rho$
inside of the system,
which is a characteristic
feature of dense systems
as
    molecular systems\footnote{
    For the one-dimensional and more complicate
    dense molecular (e.g., liquid-drop) systems,
    van der Waals (vdW) suggested the phenomenological
    capillary theory\cite{RW82} which predicted the results
    for the particle number density
$\rho$ and surface tension coefficients $\sigma$. These results are
similar to those obtained later in Refs.~\citen{wilets,strtyap}; see also
comments below.},
liquid drops, nuclei,
 and presumably, NSs. 
The realistic energy-density distribution is minimal at a certain saturation
density 
of particles (nucleons, neutrons, or nuclei)
corresponding approximately to the infinite matter\cite{bete}. As a result, 
relatively narrow edge region exists in finite nuclei or NS (crust)
in which the 
density drops sharply from its almost central value to zero. We assume
here that the part 
inside of the system far from the ES can be  changed
a little (saturation property of the dense system 
  as  hydrostatic liquid drop, nucleus
or NS in the final evolution state).

The coordinate system related to the effective
surface is defined in such 
a way that one of the spatial coordinates ($\xi$) is the distance from the 
given point to the 
ES; see Fig.~\ref{fig1},  and
\ref{appA}.
This non-linear coordinate system is
conveniently used in 
the region of nuclear and NS edges.
They  allow for an easy extraction of
relatively large 
terms in the density distribution equations for the variation 
    equilibrium
condition.
      This condition means that 
      the variation of
      the total energy $E$ over the
density $\rho$ is zero under the constraints which fix 
some
integrals of
motion beyond the energy $E$ by the Lagrange method. The Lagrange multipliers
  are determined by these constraints
  within the local energy-density theory, in particular,
  the extended Thomas-Fermi 
  (ETF) approach 
  from nuclear physics\cite{brguehak,brbhad}.
  Neglecting the other smaller-order perpendicular- and all
      parallel-to-ES
  contributions, 
sum of such terms leads to a simple one-dimensional equation (in special local 
coordinates with the coordinate normal-to-surface $\xi$);
see Fig.~\ref{fig1}.

Such an equation mainly determines approximately the density
distribution 
across the diffused surface layer of the relatively small order of 
the ratio of the diffuseness 
    parameter $a$ 
to the (mean curvature) ES radius $R$,
$a/R \ll 1$, for sufficiently
heavy systems. Notice that within this manuscript,
as in Refs.~\citen{wilets,strtyap,tyapin,strmagbr,strmagden,MS09,BM13,BM15},
the ``diffuseness parameter'', `` the crust range'', and ``the thickness of
the system edge'' have the same meaning for neutron stars.
A small parameter, $a/R$, of the expansion within the ESA  can be used
for analytical solving the variational problem
for the minimum of the system energy 
with constraints for
a fixed particle number, and other integrals of motion,
 such as angular momentum,
quadrupole deformation, etc.   
When this edge distribution of the 
density is known, the leading
static and dynamic density distributions which 
correspond to diffused surface conditions can be easily constructed. 
To do so, one has 
to determine the dynamics of the effective 
surface which is coupled to 
    the volume dynamics of the density by certain 
LDM
boundary conditions\cite{bormot,strmagbr,magstr}.
This ESA approach is
based on the catastrophe theory for solving differential equations with
a small parameter of the order of $a/R$ as the coefficient
in front of the high order derivatives in normal
to the ES direction\cite{vdW}. 
A relatively large change of the density $\rho$ on a small
distance $a$ with respect to the curvature radius $R$ takes place for the
 liquid-matter drop
(nuclei, water drops, neutron stars).
Inside of such dense systems,
the density $\rho$ is changed 
slightly around
a constant saturation density $\overline{\rho}$. Therefore,
one obtains
essential effects
of the surface capillary pressure. Another important idea of Tolman is that
we should consider as simpler as possible the neutron star 
    in terms of
the liquid
drop at a
 static equilibrium under the stability condition, i.e.,
 having a minimum of the total energy under constraints
     mentioned above.

The accuracy of the ESA was checked in 
Ref.~\citen{strmagden} for the case of nuclear physics
by comparing the results 
with the existing nuclear theories like Hartree-Fock\cite{vauthbrink} (HF)
and extended Thomas-Fermi\cite{brguehak,brbhad} approaches, based on the 
Skyrme forces 
\cite{vauthbrink,skyrme,barjac,ringshuk,blaizot,brguehak,gramvoros,krivin,CB97,CB98}, 
but for the simplest
case without spin-orbit and asymmetry terms of the energy density functional.
The direct variational principle for finding numerically 
the parameters of the tested particle density functions in simple forms
of the Woods-Saxon-like in Ref.~\citen{brbhad}
 or their powers
(Ref.~\citen{kolsan})
were applied by using the realistic Skyrme energy functional\cite{CB97,CB98}.
The main focus in Ref.~\citen{kolsan} 
aimed to the surface corrections to the
nuclear symmetry energy of spherical nuclei, see also recently published
variational ETF approach\cite{KS18,KS20}.
The extension of the ES approach
to the nuclear isotopic symmetry and spin-orbit interaction has been done in
 Refs.~\citen{MS09,BM13,BM15}. The Swiatecki derivative
terms of the symmetry energy for
heavy nuclei\cite{myswann69,myswiat85,danielewicz2,vinas1,vinas2,vinas4,vinas5,Pi09} were
taken into account  within the ESA in Ref.~\citen{BM15}. The
discussions of the progress in nuclear physics and astrophysics
within the relativistic local density approach,
can be found in reviews
Refs.~\citen{SBL23,NVR11}; see also
Refs.~\citen{CH08,Pi21,Pe23}. 
    Notice that the simplest nuclear
    Thomas-Fermi approach (without gradient terms in the energy density)
    was applied for neutron stars in Ref.~\citen{CK78}.

In the present work, we extend the ES approximation
of Refs.~\citen{strmagden,MS09,BM13,BM15}, 
to the spherical
neutron stars by the method working also for deformed systems. 
In Sect. \ref{enden}, we present the basic local
energy-density formalism
    taking into account particle density gradients which are
    responsible for surface 
     terms in finite  macroscopic dense systems.
The particle number density solutions inside of the nucleus are 
obtained analytically within the ESA expansion in 
a small leptodermic parameter $a/R$ at leading order  in 
Sect. \ref{solvolume}. The leading density distributions for the
density $\rho$ 
at the
lowest order in a relatively small diffuse surface-layer size,
$a/R \ll 1$,
are obtained analytically in Sect. \ref{densdistr0}.
The NS mass with the surface correction
is derived in Sect.~\ref{NSmass} taking into account the Schwarzschild
metric.
The surface energy 
in terms of the tension coefficients of the vdW macroscopic
capillary theory\cite{vdW,bormot,strmagbr,magstr} 
is obtained  
analytically through vdW-Skyrme forces parameters in
Sect. \ref{enertot1}. 
Equation of state in the form $P=P(\rho)$, for the system surface
    corrections, is derived through the ES at the same leading order
in Sect.~\ref{eosP}.
    The 
    TOV approach is presented in Sect.~\ref{tov} 
        for a step-like particle number
    density.
        The results of our
        calculations and comparison of the
          macroscopic and some semi-microscopic
          calculations of the EoS are discussed in Sect.~\ref{discres}.
        The main results
    are summarized in Sect.~\ref{concl}.
Some details of the
mathematical textbook relations will be shown in
 \ref{appA}.

\section{LOCAL ENERGY DENSITY AND CONSTRAINTS }
\l{enden}

The total energy $E$ for static problems 
can be written
as 
\be\l{energytot}
E=\int \d \mathcal{V}\; \epsi [\rho({\bf r})],
\ee
where  ${\cal E}(\rho)$ is the energy density\cite{strmagden,BM15},
\be\l{enerden}
\epsi\left(\rho\right) =\mathcal{A}(\rho)
+\mathcal{B}(\rho)\left(\nabla \rho\right)^2. 
\ee
The integration is carried out over the volume of 
  a  system,
  $\d \mathcal{V}=e^{\lambda/2} \d {\bf r}$,
  where $\lambda$ is the coefficient of
    the Schwarzchild metric (\ref{Schwarzgen}) for a
spherical system. As shown in Ref.~\citen{LLv2},
the multiplier $e^{\lambda/2}$ takes into account the gravitational defect of
the NS mass.
In Eq.~(\ref{enerden}), $\mathcal{A}(\rho)$
and $\mathcal{B}(\rho)$ are smooth functions of the density $\rho$ which
are coefficients in expansion of the energy density over gradients of
$\rho$.
A non-gradient part $\mathcal{A}(\rho)$ of the energy density
$\epsi$ 
 can be written as
\be\l{vol0}
\mathcal{A}=- b^{}_V\rho + \eps(\rho)+m \rho \Phi(\rho),
\ee
where $b^{}_V$ is the non-gravitational energy component for the
separation of particle from the matter, $m$ is the particle mass,
    and $\Phi$ is the macroscopic gravitational potential determined
    in more details below. The second
and third terms
take into
account the non-gravitational (like nuclear)
and gravitational contributions into the incompressibility. 
        In order to define
    properly other quantities in Eq.~(\ref{vol0}),
    $\eps(\rho)$ and $\Phi(\rho)$,
 we now use the condition for a minimum of the
energy per particle, $\mathcal{W}$, at a stable equilibrium,
\be\l{satcond}
\left(\frac{\d \mathcal{W}}{\d \rho}\right)_{\rho=\overline{\rho}} =0~,
\quad \mathcal{W}=\frac{\epsi}{\rho}~.
\ee
Near the saturation value,
$\rho \rightarrow \overline{\rho}$, one has $\mathcal{W}=\mathcal{A}/\rho$.
Therefore, there is no linear terms in expansion of $\mathcal{A}$,
Eq.~(\ref{vol0}), over powers of the difference
$\rho-\overline{\rho}$ near the saturation value
for an isolated system including the gravitation.
For instance, for $\eps(\rho)$ in Eq.~(\ref{vol0}),
one can write, generally speaking, for
    the considered dense
system (similarly as in Refs.~\citen{strmagden,MS09}),
\be\l{eps}
\eps(\rho)=\frac{K}{18 \overline{\rho}^2}~
\rho\left(\rho-\overline{\rho}\right)^2, 
\ee
where $K$ is the non-gravitational part of the
incompressibility modulus, e.g., due
to the Skyrme nuclear interaction (for nuclear
matter $K\sim 200$ MeV).
In Eq.~(\ref{vol0}), as mentioned above,
    $\Phi(\rho)$ is the main statistically (macroscopically)
averaged part of the gravitational potential.
In what follows, we may restrict ourselves by only quadratic terms
    in expansion of the smooth
function $\mathcal{A}(\rho)$ due to the
saturation property of the NS inside of the dense finite drop of the matter.

As well known\cite{RT87,LLv2}, the Einstein-Gilbert GRT equations
in the non-relativistic limit and/or
for a week gravitational field
can be transformed to the continuity Poisson equation for the
gravitational
potential $\Phi$.
As shown
in Ref.~\citen{LLv2}, in this limit, $g^{}_{00}=e^\nu$ of 
a more general Schwarzschild metric, Eq.~(\ref{Schwarzgen}), can be presented
as $e^\nu \approx 1+\nu = 1+2\Phi/c^2$, where $\Phi$ is the solution of the
Poisson equation\cite{LLv2,BK11}
$\nu=2\Phi/c^2$. 
We will use a more general
definition for the gravitational potential,
$\Phi=(c^2/2)\ln g^{}_{00}=(c^2/2) \nu$.
Instead of the Newtonian limit expansion, for the statistically
averaged potential
$\Phi$, we will use another macroscopic approximation
within the leptodermic
approach.
 Therefore, after such a 
    macroscopic averaging, $\Phi$ can be
considered as function
of the radial coordinate $r$
through the particle number density $\rho=\rho(r)$
in the form of 
expansion of $\Phi(\rho)$ over powers of $\rho-\overline{\rho}$
near
the saturation density $\overline{\rho}$ 
    for leptodermic systems.
        This is similar to the
        macroscopic Coulomb potential in nuclear physics, considered
        up to the residual
    inter-particle interaction\cite{strtyap,tyapin,MS09} but
    accounting now
    for second order terms in $\rho-\overline{\rho}$.
Up to second order, one can
take into account the 
macroscopically averaged
gravitational contribution
to
the incompressibility modulus $K$, 
\be\l{phiexp}
\Phi=\Phi_0+\Phi_1(\rho-\overline{\rho}) + (1/2)
\Phi_2(\rho-\overline{\rho})^2,
\ee
where $\Phi_0=\Phi(\overline{\rho})$,
    $\Phi_n=\partial^n \Phi(\overline{\rho})/\partial \rho^n$ (n=1,2)
are the 
derivatives of the gravitational potential
$\Phi(\rho)$ at the saturation density $\overline{\rho}$.
    According to 
    Eq.~(\ref{vol0})
    for the non-gradient part $\mathcal{A}$ of the energy density $\epsi$,
    and 
    the saturation condition (\ref{satcond}),
        one
        finds 
        for the zero constant $\Phi_1$ in the
        expansion  (\ref{phiexp}) of the gravitational
        potential $\Phi(\rho)$.
With this condition and the expansion (\ref{phiexp}),
one obtains
\be\l{phiexp2}
\Phi=\Phi_0 + (1/2)
\Phi_2(\rho-\overline{\rho})^2~.
\ee
Therefore, for $\mathcal{A}(\rho)$,
    Eq.~(\ref{vol0}), we arrive at
\be\l{vol}
\mathcal{A}=- b^{(G)}_V  \rho 
+\eps^{}_G(\rho)~,\quad b^{(G)}_V=b^{}_V - m \Phi_0~.
\ee
    It was convenient to introduce the two new quantities,
    $b^{(G)}_V$ is the total
    separation energy per particle,
    and  $\eps^{}_G(\rho)$ is the total second-order
energy density 
    component, both
modified
by the gravitational field,  
\be\l{epsKG}
\eps^{}_G(\rho)=\eps(\rho)+\frac{m}{2}
\Phi_2(\rho-\overline{\rho})^2
=\frac{K_G}{18 \overline{\rho}^2}~
\rho\left(\rho-\overline{\rho}\right)^2~. 
\ee
Here, $K_G$ is the total incompressibility modulus modified by the
gravitational field ($K_G>0$),
\be\l{KG}
K_G=K+ 9 m \overline{\rho}^2\Phi_2~. 
\ee
    Notice that our second order approximation for the gravitational potential
    $\Phi$ expansion, Eq.~(\ref{phiexp2}), 
    agrees with
    the energy density presentation up to second order
    gradients, Eq.~(\ref{enerden}).
    The latter is, in turn, rather general
    for all known Skyrme forces in nuclear physics taking
    into account the volume and surface terms.
    Thus, we 
    account for the contribution of
    a strong gravitational field in terms of separation energy $b^{(G)}_V$,
    Eq.~(\ref{vol}),
    and the total
    incompressibility $K_G$, Eq.~(\ref{KG}), 
        which will be agreed with the Schwarzschild
        metric space curvature 
            within the ES approximation.
        As we will see, this second-order
potential term is consistently related with the density
gradient squared
term of
the energy density $\epsi(\rho)$, Eq.~(\ref{enerden}), by the
equilibrium equation at leading order of
the leptodermic parameter  $a/R$.

The coefficient $\mathcal{B}(\rho)$ in front of the gradient squared term
in Eq.~(\ref{enerden}) is
given by 
\be\l{surf}
\mathcal{B}(\rho)=\mathcal{C} +\mathcal{D}\rho +
\frac{\Gamma}{\rho}~.
\ee
These terms are associated with the nuclear Skyrme interaction.
First term is related to the interaction term which is a main
reason of the
diffuse surface thickness for a non-rotating NS. 
    In this sense it has a
more general meaning including the main effective interaction in
a dense molecular system, studied by van der
Waals\cite{RW82} (vdW); see also the footnote on page 4. Therefore, we will
call this component as the vdW-Skyrme interaction.
The second term is coming from the spin-orbit interaction.  
  This interaction
might be important,
for instance, for any dense rotating liquid-drop systems with a
sharp surface edge, i.e., with
large density-gradient terms in the surface region for  a finite
leptodermic systems; see also the same general arguments for using the ETF
approach in Refs. \citen{brbhad,brguehak}.
The last term is a gradient correction to the
kinetic energy ($\hbar^2$ correction of the nuclear kinetic energy in the ETF
approach\cite{brguehak,brbhad}). It is introduced
here for a comparison of the dense and
gas systems. Thus, the forms (\ref{vol}) for $\mathcal{A}$ and
(\ref{surf}) for $\mathcal{B}$ are rather general among the simplest
analytical solutions for dense finite systems.

We should add the constraint for the variational procedure to get equations
for the static equilibrium. For non-rotated one-component 
NS system
we
have to fix the particle number $N$,
\be\l{part}
N=\int \d \mathcal{V} \rho({\bf r})~.
\ee
The integration is carried out over the volume occupied by the
gravitational mass and determined by the GRT metric.
Therefore, introducing the chemical potential $\mu$ as the Lagrange
multiplier, 
    one obtains from Eqs.~(\ref{energytot}) and (\ref{enerden})
the equation for the equilibrium,
\be\l{eq}
\hspace{-0.5cm}\frac{\delta \epsi}{\delta \rho}\equiv
\frac{\partial\mathcal{A}}{\partial\rho}
- \frac{\partial\mathcal{B}}{\partial \rho}\left(\nabla \rho\right)^2
-2 \mathcal{B}\Delta \rho =\mu~,
\ee
where $\mathcal{A}(\rho)$ and $\mathcal{B}(\rho)$ are
    given by Eqs.~(\ref{vol})
and (\ref{surf}), respectively.
Equation (\ref{part})
determines the chemical potential $\mu$ in terms of the
particle number $N$.  
For nuclear
liquid drop, one has two equations related to the two constraints for
the
fixed
neutron and proton numbers
 (Refs.~\citen{MS09,BM13,BM15}). 
They determine two (neutron and
proton) chemical potentials.
The Coulomb interaction can be taken into account for a proton
part of the
nucleus through the Coulomb potential; see Refs.~\citen{tyapin,MS09}.
The terms
like $ \propto \Delta \rho$ with constant of the proportionality are omitted
because 
they do not affect the equilibrium density because of their disappearance
 from the variational equations for the total 
 energy owing to the well-known Ostragradsky-Gauss theorem. High order
  derivative  terms
 in the energy per particle $\mathcal{E}/\rho $
  [see   Eq.~(\ref{enerden}) for $\mathcal{E} $],
 are neglected for
 simplicity in analogy of the GRT, Ref.~\citen{RT87}.
 Equation (\ref{enerden}) for  $\mathcal{E} $
 overlaps most of the Skyrme forces\cite{CB97,CB98}. As function of a 
local particle density $\rho$, the $\mathcal{E}(\rho)$ corresponds to a 
saturation condition, Eq.~(\ref{satcond}).
The
    most remarkable in this form is that the
    energy density $\epsi(\rho)$,
Eq.~(\ref{enerden}), is a sum of 
    the 
three
 terms related to the incompressible 
 constant, $-b^{(G)}_{V}$, Eq.~(\ref{vol}), and the compressible energy,
 $\propto (\rho-\overline{\rho})^2$, both including
 the gravitational components,
 Eq.~(\ref{phiexp2}), and the surface gradient terms.
 In nuclear physics, one has, instead
 of the gravitational term but similarly,
 the Coulomb potential (even without quadratic terms).
 Similarly, the  symmetry energy, and surface ($\nabla^2$) terms
 can be taken into account in nuclear
 physics and astrophysics.
Within the nuclear ETF approach, the terms being proportional to
$\Gamma$ of the gradient 
part in 
(\ref{enerden}) comes
from the $\hbar^2$ correction to the TF kinetic 
energy density\cite{brbhad}, $\Gamma=\hbar^2/18 m$, where $m$ is the
nucleon mass.
In the gradient squared part, Eq.~(\ref{surf}), 
    the $\mathcal{C}$ constant term is
typical for the potential component 
of the Skyrme energy density functional and $\mathcal{D}$ is
the constant of the
spin-orbit term\cite{brguehak} 
    in the ETF model in nuclear physics.
For a nucleus, the spin-orbit coefficient $\mathcal{D}$ 
is given by
$\mathcal{D} = -(9m/16 \hbar^2) W_0^2 $,
where  $W_0=100 - 130$ MeV fm$^{5}$ is the nuclear spin-orbit constant;
see Refs.~\citen{brguehak,CB97,CB98}.
The Coulomb part of the nuclear energy density (\ref{enerden}) is considered
similarly
as suggested in
Refs.~\citen{strtyap,tyapin}. Meaning of all terms in the energy 
density (\ref{enerden}) will be specified more below.

For the spherical and deformed system of $N$ particles, 
we may find the equilibrium
particle density
$\rho $ from the variational problem for the energy functional
(\ref{energytot}) with respect to the variations of the 
 density $\delta \rho$, 
 which obey the constraints 
 in the form: 
\be\l{constraintsNQ} 
N = \int \d \mathcal{V}\; \rho({\bf r}),\quad 
Q = \int \d \mathcal{V}\; \rho({\bf r})\; q({\bf r})~.
\ee
The last constraint  
fixes a certain deformation parameter.
The function $q({\bf r})$, for instance, can be 
a multipole moment, or it can be chosen in such a way
that Q determines
the distance between the centers of masses of the two halves of the 
stretched nucleus for fission problems\cite{strlyaschpop}. 
Thus, for the energy functional (\ref{energytot}), (\ref{enerden}) and the
constraints (\ref{constraintsNQ}), one finds the same
Lagrange variational equation (\ref{eq}) but $\mu$ would 
    be equal to
the sum
of the chemical potential and $q\mu^{}_Q$,
$\mu \Rightarrow \mu+q \mu^{}_Q $.
Lagrange multipliers $\mu$, and $\mu^{}_Q$ 
are determined by the two constraints (\ref{constraintsNQ}). 
Formally, we may consider the same
equation (\ref{eq}) for spherical and deformed systems taking into account
$q$ in $\mu$.

\section{VARIATIONAL EQUATION IN THE SYSTEM VOLUME}
\l{solvolume}

In the system volume, the terms of Eq.~(\ref{eq}) containing 
derivatives of $\rho$ are small. 
These derivatives in a normal direction become large near the nuclear edge.
For a rather wide class of deformed shapes of the dense finite system,
as well as for
the axially-symmetric, in particular, spherical one, we may assume that the 
thickness $a$ of a  system edge 
is small as compared with its mean 
curvature radius $R$,
considering $a/R$ as a small parameter. In this respect, we
define the effective surface as the points of maximum of
the gradient of a particle density $\nabla \rho$, 
as shown in Fig.~\ref{fig1} by dots in the right panel.
Locally near a point of the ES, one may 
introduce the local coordinate system $\xi, \eta$, where $\xi $ is normal
to the ES, and $\eta$ presents  two other orthogonal coordinates; see
for instance, as
shown in Fig.~\ref{fig1} 
and \ref{appA}.  In particular,
for the spherical coordinates, $\eta$ can be two spherical
angle coordinates. 
See \ref{appA},  
also for simple geometric relations for the non-linear $\xi,\eta$ coordinates.
It naturally appears the mean curvature $H$ in terms of a small
leptodermic parameter
of the ES approximation, $aH$;
see Eq.~(\ref{curvature}).
For heavy enough nuclei
near the spherical shape with the
effective radius
$R=r_0 A^{1/3}$,  one has 
    $aH = a/R \sim A^{-1/3} \ll 1$ because 
$a/r_0 \sim 1$ for realistic nuclear 
parameters, $r_0 = (4 \pi \overline{\rho}/3)^{-1/3} \approx 1.14$ fm at
the density of infinite nuclear matter,  $\overline{\rho}=0.16$ fm$^{-3}$,
and the typical diffuseness parameter $a \approx 0.8$ fm, see below 
more precise definition for the diffuseness parameter through the decrement 
of decreasing of the exponential asymptotes of particle density.
For NSs, it is well known that the typical surface diffuseness
(the NS 
 crust)
$a \approx 1$ km, and the mean effective radius $R\approx 10$ km; see, e.g.,
 Ref.~\citen{HPY07}.
Therefore, the ratio $a/R$
can be also considered as a small parameter, and the leptodermic approximation,
$a/R \ll 1$, can be used too. 

The largest terms in Eq.~(\ref{eq}) within the region of 
a sharp density descent are 
the second-order derivative of particle density $\rho$ in the 
$\xi$ direction, normal to the ES, $\d^2\rho/\d \xi^2$, and its
derivative square, $(\d\rho/\d \xi)^2$, both of the order of 
$(\overline{\rho}/a)^2 \propto (R/a)^2$; see the expression for the Laplacian
and gradient in $\xi,\eta$ coordinates in \ref{appA}. 
Our approach is based on expansion in the same 
small parameter $aH$ (for a nucleus, $aH = a/R\sim 1/k_FR \sim A^{-1/3}$,
where $k^{}_F$ is the Fermi momentum in units of $\hbar$).
 This 
leptodermic 
approach is used in the liquid-drop 
and the extended Thomas-Fermi 
approach. 
Following
 Refs.~\citen{strtyap,strmagbr,strmagden,MS09,BM13,BM15},
 we shall call this statistical\footnote{The ETF in nuclear physics
 is the statistical
semiclassical approach because for convergence of the expansion in $\hbar$ of
the partition function 
we have first to average statistically it removing strong (e.g., shell)
oscillations, and thus, get a smooth behavior of the
partition function\cite{brbhad}.}
semiclassical ETF approach as the ES
approximation. 
We have to evaluate also the derivatives of the particle density
and gravitational terms in the energy density $\epsi(\rho)$,
Eq.~(\ref{enerden}), and Lagrange equation (\ref{eq}).
For simplicity, the gravitational terms [see Eqs.(\ref{phiexp2})
   and 
  (\ref{eq})] are assumed to be of the same order as
other non-gradient terms. 
    Notice that according to these estimations,
    the transformation of radial derivatives
due to the Schwarzschild radial curvature can be taken into account
through the leptodermic parameter $a/R$ because they are important
near the ES. The gravitational corrections
are in contrast
to those due to the Coulomb potential for the
nuclear liquid drop, where the Coulomb corrections to a constant are
assumed to
be negligible
as $a/R$. The latter corrections\cite{tyapin} are 
of the relative order
$\sim (e^2/\hbar c)(\hbar c/b^{}_{V} r_0) (Z^2/A^{4/3})/2 \siml 0.2$.

As function $\epsi(\rho) $ obeys the saturation property (\ref{satcond}),
in the system volume at $\xi \siml \xi_{vol}$
($\xi <0$ and $|\xi| \gg a$), where
$\rho-\overline{\rho}$ is small
asymptotically far from the ES, one can expand the non-gradient
function $\mathcal{A}$ of the energy density $\epsi(\rho)$,
Eq.~(\ref{enerden}), up to second order 
including the
compression term of the second order in powers of
    $\rho-\overline{\rho}$, $\mathcal{A}=\mathcal{A}_V$; see
    Eqs.~(\ref{vol}), (\ref{epsKG}), and (\ref{eps}),
\be\l{enerdenvol}
\epsi(\rho) \rightarrow \mathcal{A}_{V} =
-b^{(G)}_{V}\rho +
\frac{K_{G}}{18\overline{\rho}^2}~\rho(\rho-\overline{\rho})^2~.
\ee
The zero and second order terms of expansion of the gravitational
potential, Eq.~(\ref{phiexp2}), are
included in
the zero and second order
terms through constants $b^{(G)}_V$ [Eq.~(\ref{vol})] and $K_G$
[Eq.~(\ref{KG})], respectively.

Introducing
the dimensionless quantities for convenience to exclude the
transformation of units,
\be\l{units}
y=\frac{\rho}{\overline{\rho}}, \quad x=\frac{\xi}{a}~,
\ee
one can present Eq.~(\ref{enerdenvol})
in the following way:
\be\l{varepsvol}
\eps^{}_G(\rho)\equiv
\mathcal{A}_V(\rho)+ 
b^{(G)}_V\rho =
\frac{K_{G}\overline{\rho}}{18}~\epsilon(y),
\ee
where
\be\l{epsilon}
\epsilon(y)=y(1-y)^2~.
\ee
Discarded terms are
of the order of $((\rho-\overline{\rho})/\overline{\rho})^3 \sim (a/R)^3$.
Neglecting gradient terms in Eq.~(\ref{eq})  
 in the system volume
 and using the approximation (\ref{phiexp2})  
 for the gravitational potential $\Phi$,
 for simplicity,
     up to quadratic terms over $\rho-\overline{\rho}$,
one can reduce equation (\ref{eq})  as 
\be\l{eqvol}
\frac{\partial \eps^{}_{G}}{\partial \rho}\equiv
\frac{K_{G}}{9}~(\rho-\overline{\rho})=\mathcal{M}~,
\ee
where $\mathcal{M}$ is an integration constant.
Solving this equation with respect to the particle number density $\rho$,
one obtains
\be\l{solvol}
\rho=\overline{\rho}\left(1 + \frac{9\mathcal{M}}{K_{G}}\right),\quad
\mathcal{M}=\mu + 
b^{(G)}_{V}~.
\ee
Therefore,  $\mathcal{M}$ is the surface (capillary) correction
to the leading component
of the
chemical potential $\mu$ [introduced above as the Lagrange multiplier,
determined through the constraint
  (\ref{part})]. 
As seen from Eq.~(\ref{solvol}),
one finds the relatively small surface 
    correction,
$9\mathcal{M} /K_G \sim a/R$, to the saturation constant value
$\overline{\rho}$; see Refs.~
\citen{strtyap,tyapin,strmagbr,MS09,BM13,BM15}. Therefore,
$\mathcal{M}$ is also the capillary pressure
correction; see also Ref.~\citen{vdW}.

\section{DENSITY NEAR
  THE EFFECTIVE SURFACE}
\label{densdistr0}

In this section, we will  analytically  solve
equation (\ref{eq}) for
the distribution of the 
particle density $\rho$ through the ES at leading order in small 
parameter $a/R$. Equation (\ref{eq}) is 
a typical 
differential equation in
the catastrophe theory. 
 In such differential equations, a small
coefficient (of the order of $a/R$ in our case) appears in front of
the highest  order derivative. 
 These terms become important because of
the product of this small coefficient (of the order of  $a/R$) and
large (of the order of $R/a$ and high) 
 can be of zero order 
in this leptodermic
parameter. At the leading order, we need to 
keep only leading terms in small parameter $a/R$ in the Lagrange equation
(\ref{eq}). 
These are the second order derivatives $\rho''$,  and
first-order derivative squares $(\rho')^2$, 
 over their 
$\xi$ variable, 
 according to Eqs.~(\ref{laplacian})
 and (\ref{metrictensor}), $\rho'=\partial \rho/\partial \xi$.
  We may develop some iteration procedure to find the solutions
 with improved precision in $a/R$ in terms of the solutions of the
 leading order. As mentioned above,
    the transformation of radial derivatives
    due to a smooth Schwarzschild radial curvature 
      for the spherical case
    can be taken into account
near the ES through the re-definition of the leptodermic parameter $a/R$.

Multiplying Eq.~(\ref{eq}) by the 
derivative 
$\partial \rho/\partial \xi$, 
    from Eqs.~(\ref{eq}), (\ref{vol}),
and (\ref{surf})
at leading order, 
one obtains 
\be\l{eq2}
\frac{\d }{\d \xi} 
\left[\left(\mathcal{C} +\mathcal{D} \rho +\frac{\Gamma}{4 \rho}\right)\;
\left(\frac{\d \rho}{\d \xi}\right)^2 -\eps^{}_G\left(\rho\right)
\right]=0~,
\ee
 where 
\be\l{varepsilon}
\varepsilon^{}_{G}(\rho)=\frac{\overline{\rho}K_{G}}{18}\epsilon(y),
\ee
and $\epsilon(y)$ is a given function of $y$, which is
determined by 
a short- and long-range inter-particle interaction.
In particular, one can use a simple
quadratic approximation (\ref{epsilon}) 
    far from the critical point
where $K_G=0$.
    In Eq.~(\ref{eq2}), the term 
 with the 
coefficient $\mathcal{C}$ in parentheses in
front of the derivative squared
of the density $\rho $ is the main term, e.g.,
as related to the vdW capillary theory.
In particular, it is the main term, associated with the Skyrme
interaction, with respect to the spin-orbit term $\mathcal{D}\rho$ and,
moreover, the gradient correction $\Gamma/(4\rho)$ to the
kinetic energy of the system in nuclear physics. Let us first
assume the same for the NS liquid drop. Therefore, $a$ should be
determined through the constant $\mathcal{C}$.
Otherwise, we have to include average of whole coefficient in front of the
derivative square term in Eq.~(\ref{eq2}).

With employing the boundary conditions $\rho \to 0$ and $\rho' \to 0$
for $\xi \to \infty$, one can easily
integrate equation (\ref{eq2}).
Finally, we come to a simple ordinary
1st order differential
equation for $\rho $, 
depending only on the normal-to-ES 
coordinate $\xi$, at leading order in a small
parameter $a/R$, that is
\be\l{eq0}
\frac{\d \rho}{\d \xi}=-\sqrt{\frac{ \eps^{}_{G}(\rho)}{\mathcal{C} +
    \mathcal{D} \rho +\Gamma/4\rho}}~.
\ee
Introducing several dimensionless quantities,
\be\l{yxedef}
y=\frac{\rho}{\overline{\rho}}, \quad x=\frac{\xi}{a},\quad
\epsilon = 
\frac{18\eps^{}_{G}}{K_{G}\overline{\rho}}~,
\ee
one can 
    relate the crust thickness $a$ to the vdW
interaction constant $\mathcal{C}$ by
\be\l{adef}
a=\sqrt{18\mathcal{C}\frac{\overline{\rho}}{K_G}}~.
\ee
Using these definitions,
one can present Eq.~(\ref{eq0}) in  the 
following dimensionless form:
\be\l{eq0yx} 
\frac{\d y}{\d x} =-\sqrt{\frac{y\epsilon(y)}{y + \beta y^2 + \gamma}}~,
\ee
 where
\be\l{consts}
\beta=\frac{\mathcal{D}\; \overline{\rho}}{\mathcal{C}}, \quad 
\gamma=\frac{\Gamma}{4 \overline{\rho}\; \mathcal{C}}~. 
\ee
From the asymptotes of the
explicit analytical expressions for the particle density
$y(x)$ at large $x $, one may see 
a typical behavior, 
$y(x) \propto e^{-x}$, where $x=\xi/a$,
that is a reason to call $a$, Eq.~(\ref{adef}), as the 
crust  diffuseness parameter.
Thus, we may check
that $a$ has exactly the same meaning as that
introduced above in terms 
of the small parameter $a/R$.  As noted above, 
in the derivation of Eq.~(\ref{eq0yx}) within leading
order of the ES expansion, we
reduced the second order differential equation (\ref{eq}) to
the first order one by integrating with using the standard 
boundary conditions $y \to 0$ and $\partial y/\partial \xi \to 0$ and
 $y \epsilon(y) \to 0$ for the definition of the $\epsilon(y)$ asymptotes.
Finally, all parameters of the energy density in 
Eq.~(\ref{enerden}) are reduced within this leading order ES 
approximation to the only two dimensionless constants;
see Eq.~(\ref{consts}).
For nuclear physics,
\be\l{constsNP}
\beta=
-\frac{27 W_0^2}{128 \pi \;k_F\; r_0^6\; \eps_F \;{\cal C}},\quad 
\gamma=\frac{\pi \eps^{}_{F}\; r_0^5}{27 {\cal C} 
\left(k^{}_{F}\; r^{}_{0}\right)^2}, 
\ee
where $\eps^{}_{F}=\hbar^2 k^{2}_{F}/2 m \approx 37$ MeV is the Fermi energy for 
the Fermi momentum, $k_F=(3 \pi^2\; \overline{\rho}/2)^{1/3}$ 
(in units of $\hbar$).
We are coming to the realistic nuclear data for 
the infinite-matter particle density, $\overline{\rho}=0.16$ fm$^{-3}$.

Single boundary condition which we need for the unique solution of 
the first order differential equation can be easily
found 
within the leading
ES approximation from the definition
of the ES as the points of maximal values of the derivative, 
$\partial \rho/\partial \xi$ at $\xi=0$, namely, $y''(0)=0$ at $x=0$. 
Differentiating equation (\ref{eq0yx}) over $x$, one
obtains the algebraic
equation for the position of the ES, $y=y_0$ at $x=0$,
\be\l{boundcond}
(\gamma-\beta y^2)\epsilon(y_0) +
y_0(y_0+\beta y_0^2+\gamma)\frac{\d \epsilon(y^{}_{0})}{\d y}=0~, 
\ee
defined by function $\epsilon(y)$; see, e.g., Eq.~(\ref{epsilon}).

For any function $\epsilon(y)$,  Eq.~(\ref{eq0yx}) can be
easily integrated.
The integration constant is determined by the boundary 
condition (\ref{boundcond}).
The leading-order solution of Eq.~(\ref{eq0yx})
can be found
explicitly  
in the inverse form:
\be\l{sol0xy}
x=-\int_{y_0}^{y}\d w \sqrt{\frac{w +\beta w^2 
    + \gamma}{w\epsilon(w)}}~.
\ee
For the expression (\ref{epsilon}) for $\epsilon(y)$,
this integral can be calculated analytically in terms of the elementary
functions for any parameters $\beta$ and $\gamma$ within the condition
$a/R \ll 1$.
Note that the solution (\ref{sol0xy}) to equation
(\ref{eq0yx}) is considered
at leading order in small parameter 
$a/R$. This solution satisfies asymptotically 
the condition of its matching with the volume 
    result, Eq.~(\ref{solvol}),
of the same order at the point  $x$ corresponding 
$\xi \approx \xi_{vol}$ defined in Section \ref{solvolume}, 
$y \to 1 $ exponentially  
for $x \to -\infty$.

Let us consider the examples of simple solutions $x=x(y)$, Eq.~(\ref{sol0xy}),
of Eq.~(\ref{eq0yx}) with the boundary condition, Eq.~(\ref{boundcond}),
for the quadratic expression
$\epsilon(y)=y(1-y)^2$. 

(i) For $\gamma=0$ and $\beta=0$,  neglecting
the gradient correction to the kinetic energy 
    density ($\Gamma=0$ in
  Eq.~(\ref{surf})), and spin-orbit ($\mathcal{D}=0$ in Eq.(\ref{surf}))
terms, one finally obtains\cite{strmagden,MS09}
\be\l{b0g0}
y(x)=\tanh^2[(x-x_0)/2],\quad
x<x_0=2 {\rm arctanh}(1/\sqrt{3})~.
\ee
This solution is related to the gradient squared term due to
the main nuclear Skyrme, or molecular van der Waals
interaction
term [$\mathcal{C} \neq 0$ in Eq.(\ref{surf})].
It is very asymmetric with respect to the ES value,
$y^{}_0=y(0)=1/3$;
see Fig.~\ref{fig1}(b). 

 (ii) Keeping only $\Gamma\neq 0$ term in Eq.~(\ref{eq0}),
 but neglected gradient terms of the
 vdW-Skyrme
 ($\mathcal{C}=0$)
 and spin-orbit ($\mathcal{D}=0$) interaction, one
  obtains\footnote{Dimensionless equation in this case is
  $\d y/\d x=-(y \epsilon(y))^{1/2}$ for $a=[9\Gamma/(2 K_G)]^{1/2}$. Equation
  for the ES, $x=0$, $\epsilon(y_0)+y_0 \d \epsilon(y_0)/ \d y=0$, has
  the solution $y_0=1/2$. Integrating analytically the equation for $y(x)$
  in this case, one obtains the expression (\ref{wilets}).}
  the Wilets solution\cite{wilets}, derived early for the 
      semi-infinite
  system,
  \be\l{wilets}
y(x)=1/(1+e^x)~.
  \ee
  This solution is symmetric with respect to the ES because of
  $y(0)=1/2$. This Fermi-gas
  form
  of the solutions
  is used for the variational problems within the ETF model,
  in contrast to the
  dense liquid-drop solution (\ref{b0g0}), related to the vdW-Skyrme
  interaction;
  see Fig.~\ref{fig1}(b).

  Figure \ref{fig1}(b) shows the particle density,
  $\rho(r)=\overline{\rho} y((r-R)/a)$, as function of
  the radial coordinate $r$ for typical parameters, the thickness of the
  NS diffused layer (crust) $a=1$ km, and
  the effective radius $R=10$ km,
    in  units of the saturation value $\overline{\rho}$.
  We compare the two dimensionless solutions for the main asymmetric
  particle density $\rho(r)/\overline{\rho}$.
  One of them (solid line) is
  related to the corresponding vdW-Skyrme interaction ``1'';
  see Eq.~(\ref{b0g0}) for the
  universal dimensionless particle density of a dense liquid-drop,
  $y=\rho/\overline{\rho}$, as
  function of $x=(r-R)/a$. Another limit case corresponds to the
  symmetric Wilets solution based on
  Eq.~(\ref{wilets}) for zero constants $\mathcal{C}$
  and $\mathcal{D}$
  in Eq.~(\ref{surf}) for $\mathcal{B}$ (only the gradient correction to
  the kinetic energy is taken into account, i.e. for a gas system).
  This solution ``2'' is often used for the variational version of
  this problem. These two curves are very different in the surface layer,
  especially outside of the system far away from the NS, and almost the same
  inside of the NS. Notice that so far
the effective surface method can be applied for the deformed
NS under the condition of smallness of crust thickness $a$ over the effective
NS radius $R$ determined by the mean surface curvature $H$, $aH=a/R\ll 1$; see
\ref{appA}. However, in the following sections,
our macroscopic method will be applied for the simplest solved analytically
case of the spherical symmetry.

\section{NS mass surface correction}
\l{NSmass}

Let us derive the NS mass M within our macroscopic ESA approach,
\be\l{MNS}
M=m \int \rho \d \mathcal{V}~,
\ee
where $m$ is the particle 
(nucleon, nucleus, or molecule) mass.
The integration is carried out over the spatial volume,
\be\l{dV}
\d \mathcal{V}=J(r) r^2\d r~ \sin \theta~ \d \theta \d \varphi~,\quad
J(r)=e^{\lambda/2}~.
\ee
This element of the spatial volume is
associated with
the Schwarzschild interior
spherical coordinate system (see Refs.~\citen{RT87,LLv2}),
     \be\l{Schwarz}
        {\rm d} s^2=-\frac{{\rm d} r^2}{1-r^2/R^2_{\rm SM}}-
     r^2 {\rm d}\theta^2 -
     r^2 \sin^2\theta {\rm d} \phi^2
      + \left[A_{\rm SM} -
       B_{\rm SM}\sqrt{1- r^2/R^2_{\rm SM}}\right]^2 {\rm d} t^2~, 
     \ee
         where $R_{\rm SM}$, $A_{\rm SM}$, and $B_{\rm SM}$ are some constants; see
     Ref.~\citen{RT87}, and more details below. Here and below
     we will use the subscript SM to relate the quantities to the
     Schwarzschild metric, Eq.~(\ref{Schwarz}).
       The Schwarzschild radius $R_{\rm SM}$ is given by\cite{RT87,LLv2}
     \be\l{RTOV}
     R_{\rm SM}=\left(\frac{8\pi G\epsi_0}{3 c^4}\right)^{-1/2}, 
     \ee
     where $G$ is the gravitational constant,
     $\epsi_0=\mathcal{A}(\overline{\rho})=
     -b^{(G)}_{V}\overline{\rho}$, Eq.~(\ref{vol}). 
    Then, according to Eqs.~(\ref{dV}) and (\ref{Schwarz}),
    for the radial part of the Jacobian factor
    $J$
    in this transformation, one has
\be\l{J}
J(r)=
\left(1-\frac{r^2}{R^2_{\rm SM}}\right)^{-1/2}.
  \ee
  The Jacobian $J$ takes approximately
  into account the gravitational defect mass\cite{LLv2}.

Adding and subtracting $\overline{\rho}$ in the integrand of Eq.~(\ref{MNS}),
one can identically rewrite the NS mass $M$ in the following way:
\be\l{MNStot}
M=  M_{V} + M_{S}~.
\ee
Here, the first volume component
of the NS mass, $M_{V}$,  is given by
\be\l{MNSV}
M_{V}=4\pi m \overline{\rho} \int_0^R
\frac{r^2{\rm d} r}{\left(1-r^2/R_{\rm SM}\right)^{1/2}}
=
4\pi m \overline{\rho} R_{\rm SM}^3
f\left(\frac{R}{R_{\rm SM}}\right)~,
\ee
where
\be\l{mcor}
f(z)=\frac12 \left[\arcsin(z) -z \sqrt{1-z^2}\right]
\ee
for $0 < z < 1$.
The second term, $M_{S}$, in Eq.~(\ref{MNStot}) is the
surface component of the NS mass,
\be\l{MNSS}
M_{S}= \frac{m S}{R^2} \int (\rho-\overline{\rho}) J(r) r^2 \d r~,
\ee
where $S$ is the surface area of the spherical system,  $S=4\pi R^2$.
The integral over 
the radial coordinate $r$ is taken effectively over a small
 diffuse crust region of
the order of $a$, 
    where $\rho-\overline{\rho}$ essentially differs from zero.
A smooth 
Jacobian $J(r)$, Eq.~(\ref{J}), multiplied 
    by $r^2$, for $r$ far away from
the Schwarzschild radius $R_{\rm SM}$, can be taken
    approximately off
    the integral in Eq.~(\ref{MNSS}) over the radial
    coordinate $r$ at the ES $r=R$, $J(r) \approx J(R)$.
    Using the differential equation (\ref{eq0}) for the density
    $\rho$ at leading order in $a/R$
    in the local
coordinate system $\xi, \eta$ (\ref{appA}),
one can transform the integration variable $\xi$ to $\rho$,
$\d r=\d \xi=(\d \xi/\d \rho) \d \rho$.
    Thus, the surface correction, $M_{S}$, Eq.~(\ref{MNSS}),
        in terms of the
dimensionless density $y=\rho/\overline{\rho}$,
is resulted in 
\be\l{MNSS1}
M_{S}\approx
-m a S \overline{\rho} J(R)
\int_0^1\frac{(1-y) \sqrt{y+\beta y^2 +\gamma}}{\sqrt{y\epsilon(y)}}
 \d y,
\ee
where
$J(R)$ is given by 
Eq.~(\ref{J}) for $J(r)$ at $r=R$,
and
$\epsilon(y)$ can be parameterized by using
    the vdW-Skyrme interaction model
    but with NS parameters. 
    For the simplified (quadratic) expression of $\epsilon(y)$,
        Eq.~(\ref{epsilon}),
    the integral in Eq.~(\ref{MNSS1})
    can be taken analytically
    in terms of the elementary functions.
    In particular, for the vdW-Skyrme case (i),
$\beta=\gamma=0$, Eqs.~(\ref{b0g0}) and (\ref{J}) at $r=R$,
one explicitly finds from Eq.~(\ref{MNSS1})
\be\l{MNSSfin}
M_{S}\approx -8 \pi m
\overline{\rho} R^2 a \left(1-R^2/R^2_{\rm SM}\right)^{-1/2}. 
\ee
Finally, we arrive from Eqs.~(\ref{MNStot}), (\ref{MNSV}), and
(\ref{MNSSfin}) at the simple result
 in the case (i):
\be\l{MNSfin}
M=M_{V}
\left[1-
\frac{2a R^2}{R^3_{\rm SM}f\left(R/R_{\rm SM}\right)
\left(1-R^2/R^2_{\rm SM}\right)^{1/2}}\right],
\ee
where $M_{V}$, Eq.~(\ref{MNSV}), is the volume part of the total
NS mass $M$
[see Eq.~(\ref{MNStot})], i.e., the mass $M$ at $a=0$,
and $f(z)$ is given by Eq.~(\ref{mcor}). 

For completeness, one also can present the NS mass with
the surface correction in
    the Wilets case (ii), $\mathcal{C}=\mathcal{D}=0$,
    Eq.~(\ref{wilets}),
 \be\l{MNSW}
M=M_{V}
\left[1-
\frac{6a \sqrt{\gamma} R^2}{R^3_{\rm SM}f\left(R/R_{\rm SM}\right)
\left(1-R^2/R^2_{\rm SM}\right)^{1/2}}\right],
\ee
    where $\gamma$ is the dimensionless constant of the gradient correction
    to the
kinetic energy density; see Eq.~(\ref{consts}).

\vspace{0.8cm}
\begin{figure*}[!ht]
  \vskip1mm
  \centerline{
    \includegraphics[width=6.5cm,angle=-90]{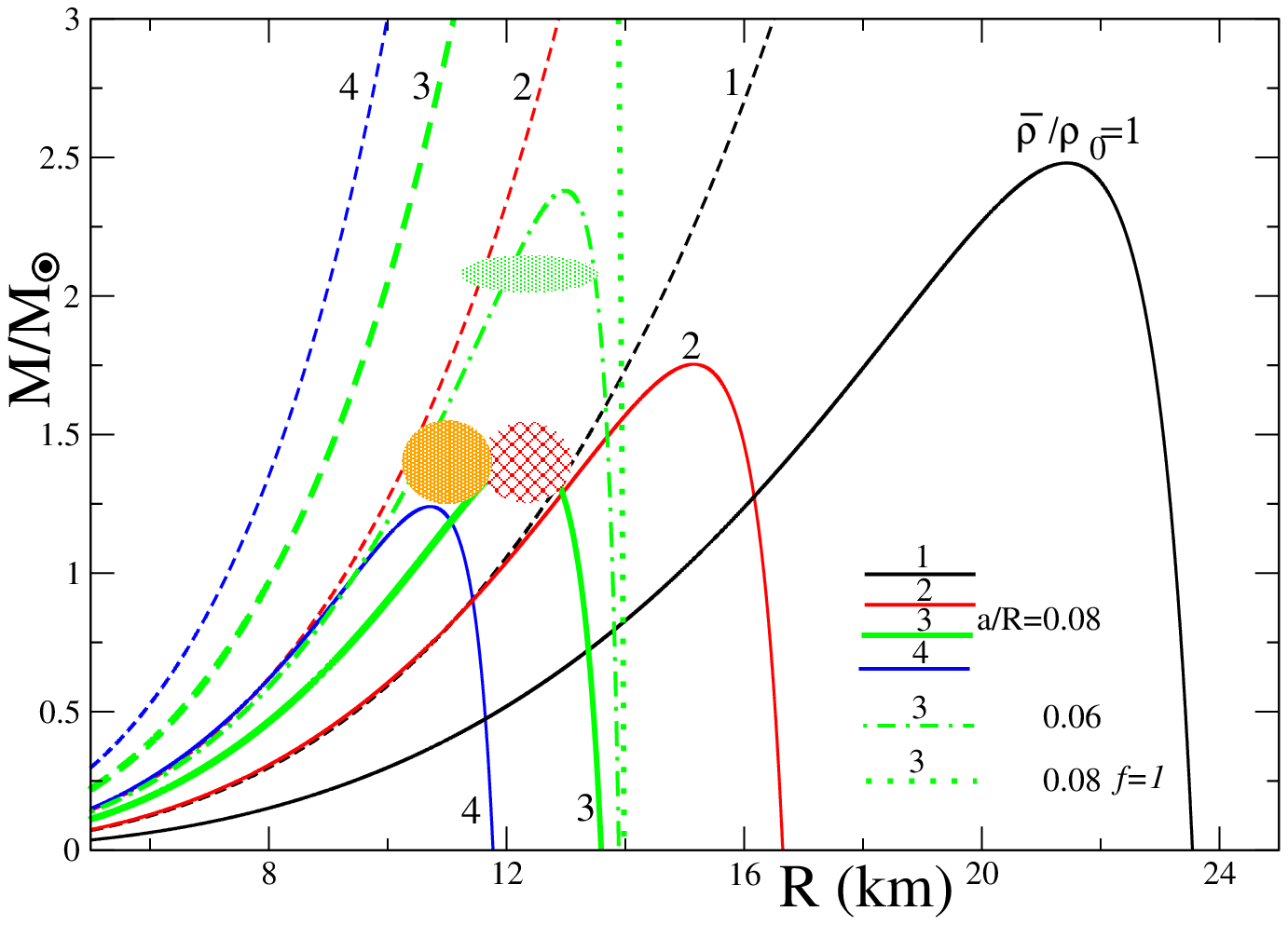}}

  \vspace{-0.8cm}
  \caption{{\small
        NS masses
        $M(R)$ [Eq.~(\ref{MNSfin})
        in Solar units $M_\odot$]
        as
         function of
        the  ES curvature radius $R$ (in km)
         are shown by black $\overline{\rho}/\rho^{}_{0}=1$,
         red 2, green 3 and blue 4 solid lines, 
        where $\rho_{0}=0.16 $ fm$^{-3}$,
         for a leptodermic parameter $a/R=0.08$.
         Similar dashed lines are the
         corresponding volume masses $M_{V}$, Eq.~(\ref{MNSV}).
                 Sensitivity of mass $M$ for the $a/R$ variation is
                 shown by the comparison of the green 
                     solid ($a/R=0.08$)
                 and dash-dotted
         ($0.06$)
         lines. The behavior of the gravitational coefficient $f$
         is seen from comparison of the green solid versus the dotted
         ($f=1$ at $\lambda=0$) curve (both at the same $a/R=0.08$).
         Red, orange, and green spots show the experimental data on
         the NS J0030+0441 (Ref.~\citen{GR21}),
         GW170817 (Ref.~\citen{CC20}), and J0740+6620
         (Ref.\citen{TR21}), respectively. 
    }}
\label{fig2}
\end{figure*}

\vspace{-0.8cm}
\begin{figure*}[!ht]
  \vskip1mm
  \centerline{\includegraphics[width=5.9cm]{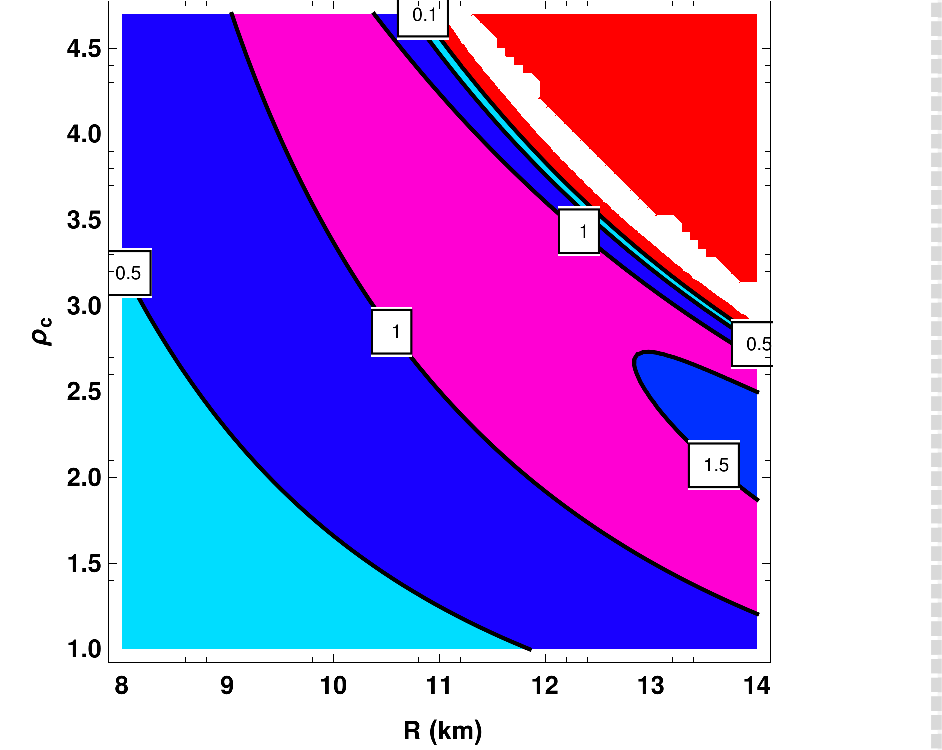}
~
  \includegraphics[width=5.9cm,angle=0]{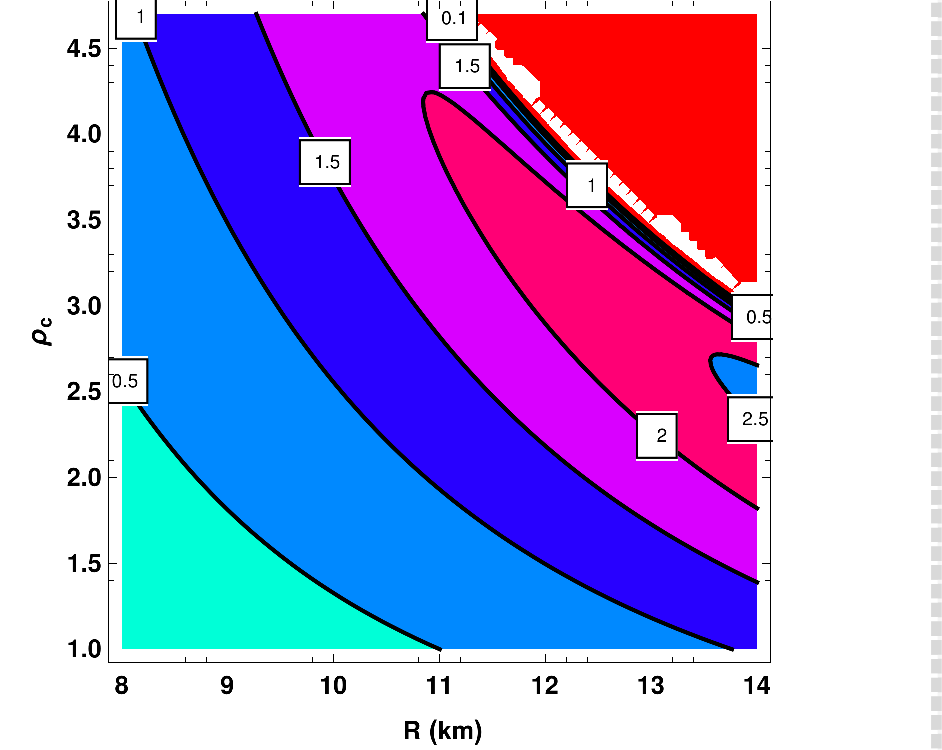}}

  \vspace{-0.4cm}
    \vskip1mm\caption{{\small
          Contour plots for the NS mass
          $M(R)$ (Eq.~(\ref{MNSfin})
          in Solar units $M_\odot$) as
         function of
         the  ES curvature radius $R$ (in km),
         and central density $\rho_c$
         (in nuclear saturation-density units
         $\rho_{0}=0.16 $ fm$^{-3}$)
         for a leptodermic parameter $a/R=0.08$ (left) and $0.06$
         (right). Red areas show non-physical regions
         $R>R_{\rm SM}$, and white ones present indeterminations
         near $R=R_{\rm SM}$.
         }}
\label{fig3}
\end{figure*}

Figure~\ref{fig2} shows the mass distribution $M(R)$,
Eq.~(\ref{MNSfin}),
as function of the
effective radius $R$
by using
only two  physical
parameters, the relative crust thickness $a/R$, 
and the central NS particle-number density $\overline{\rho}$ in units
of the nuclear matter-saturation density $\rho^{}_{0}=0.16$ fm$^{-3}$,
$\overline{\rho}/\rho^{}_{0}= 1-4$, as examples.
The surface effect is measured by the relative
difference
between the full NS mass, $M=M_{V}+M_{S}$ [see Eq.~(\ref{MNSfin})
with Eqs.~(\ref{MNSV}) and (\ref{mcor}), solid lines]
 and its
corresponding volume
component $M_{V}$ ($a=0$, dashed curves).
As seen from Fig.~\ref{fig2}, the surface component is rather
notable even at
a small leptodermic parameter $a/R=0.08$. 
    The significant
influence of the gravitational
forces through the Schwarzschild metric (\ref{Schwarz}) is shown by
comparing the solid [Eq.~(\ref{mcor}) for $f$] and dotted
($f=1$) green lines,
both taken at $a/R=0.08$. It depends also essentially
on the value of $a/R$ 
through the equilibrium condition; cf. dashed-dotted ($a=0.06$)
and solid ($a=0.08$)
green lines.
The NS mass $M(R)$ is the non-monotonic
function of $R$ for a constant central NS density,
$\overline{\rho}$,
because of the surface component, $M_{S}$, Eq.~(\ref{MNSSfin}).
This is
in contrast to the monotonic behavior of the volume mass,
$M_{V} \propto R^3$, in the Cartesian case of the small Newtonian
gravitational
limit at $a=0$; see also the dashed ($a=0$) at finite $f(R/R_{\rm SM})$
[Eq.~(\ref{mcor})] curves.
For any given value of $\overline{\rho}$,
one finds a
rather pronounced maximum in dependence of the full mass $M(R)$,
Eq.~(\ref{MNSfin}). As well known\cite{RT87,LLv2},
the physical values of the NS radius for the Schwarzschild metric,
Eq.~(\ref{Schwarz}) inside of
the system, have to obey
the condition, $r_g < R < R_{\rm SM}$. This means
on the left side of the sharp addiction decline, i.e.
more pronounceable near the
maxima in Fig.~\ref{fig2}. These maxima are increased from about 1.2
to 2.5 $M_\odot$ for decreasing $\overline{\rho}/\rho^{}_0$ from 4 to 1
at $a/R=0.08$
as example, respectively. The NS mass for each of
these curves at a given central value, $\rho_c=\overline{\rho}$,
disappears sharply 
in the limit $R \rightarrow R_{\rm SM}$,
and does not exist at $R>R_{\rm SM}$. As mentioned above, we should emphasize
that our derivations for the surface component of the NS mass, $M_{\rm S}$,
fail near the point $R=R_{\rm SM}$ because
of the obvious singularity of the Jacobian $J(r)$,
Eq.~(\ref{J}), at $ r = R = R_{\rm SM}$.
As seen from Fig.~\ref{fig2}, our results are in reasonable agreement with
the experimental data\cite{CC20,GR21,TR21};
see also Ref.~\citen{OL20} for the theoretical
results about the neutron stars. For smaller NS masses,
$M \approx(1.2-1.5)M_\odot$ for the NS pulsars GW170817 (Ref.~\citen{CC20})
and
J0030+0451 (Ref.~\citen{GR21}) with the slightly different radii,
$R=10.3-11.8$ km (orange spot) and $R=11.6-13.1$ km (red spot),
one finds respectively good agreement
with our results 
for the central density $\overline{\rho}=3 \rho^{}_{0}$,
with $a/R=0.08$. Larger mass of the NS pulsar J0740+6620 (Ref.~\citen{TR21}),
$M=(2.0-2.1) M_{\odot}$ and radius $R=11.3-13.6$ km (green spot),
corresponds to the same central density,
$\rho_c=\overline{\rho}$, but with smaller
leptodermic parameter $a/R=0.06$.

Figure~\ref{fig3}
shows contour plots for the NS mass, $M$ (in Solar units)
as functions of its radius, $R$, and central density, $\rho_c$,
($\overline{\rho}$
in units of
$\rho^{}_0=0.16$fm$^{-3}$) for the leptodermic parameter values,
$a/R=0.08$ (left) and $0.06$ (right). As seen from the left
figure, one for instance finds the line for the NS mass
$1.5 M_\odot$ for radius
$R=12.6-14.0$ km
and central density, $\overline{\rho}/\rho^{}_0=1.8-2.7$.
In the right figure we can see the line for the NS mass
$2.0 M_\odot$ for radius $R=10.9-14.0$ km
and central density, $\overline{\rho}/\rho^{}_0=1.8-4.3$.
They are close to the experimental data shown in Fig.~\ref{fig2}.
The NS mass structure is changed significantly with radius $R$ and
relative central density $\rho_c$, also with the leptodermic parameter.
The calculations become unstable with approaching the line $R=R_{\rm SM}$,
as shown by white areas. 
The physical region is below this line
because of the restriction by
the condition $r_g<R<R_{\rm SM}$.

\section{The NS surface energy}
\l{enertot1}

Let us derive now the NS total energy $E$ which is
similar to the main two (volume and surface) terms of the
Weizs\"aker mass formula, basic in
nuclear physics,
but with the analytical expression for
its tension coefficient and accounting for the gravitational forces. 
According to the basic definitions, the total energy
$E$, Eq.~(\ref{energytot}), can be presented as
\be\l{enertot}
E=\int \d \mathcal{V}  \mathcal{E}[\rho({\bf r})] \approx E_{V}+E_{S}~,
\ee
where $\d\mathcal{V}$ is given by Eqs.~(\ref{dV}) and (\ref{J}), and
$E_V$ is the volume part of the total energy. Following the mass derivations
in the previous section, one has
\be\l{enertotv}
E^{}_{V}=-b^{(G)}_{V}\overline{\rho}\int \d \mathcal{V}
=
-4 \pi b^{(G)}_{V} \overline{\rho} R_{\rm SM}^3 f\left(\frac{R}{R_{\rm SM}}\right),
\ee
where $b^{(G)}_{V}$ is given by Eq.~(\ref{vol}),
and $f(z)$ is
shown in Eq.~(\ref{mcor}).
 In Eq.~(\ref{enertot}), $E_{S}$ is the surface part,
\be\l{enertots}
E_{S} \approx \sigma S, 
\ee
 where $S=4\pi R^2$ is the surface area value, and $\sigma$
is the tension coefficient. The leading expression for $\sigma$ in our
leptodermic expansion over small parameter $a/R$ is given by\footnote{
Similar expressions as Eqs.~(\ref{sigma}) and (\ref{sigma1}) at
$\mathcal{D}=\Gamma=0$ were obtained early by van der Waals;
see Ref.~\citen{RW82}.}
\be\l{sigma}
\sigma=2 J(R)\int_{-\infty}^\infty \d \xi
\left[\mathcal{C} + \mathcal{D}\rho +\frac{\Gamma}{4\rho}
\right]
            \left(\frac{\partial \rho}{\partial \xi}\right)^2,
\ee
where $J(R)$ is the Schwarzschild metric
Jacobian $J(r)$
at $r=R$; see Eq.~(\ref{J}).
In these derivations, we present locally
the integration over $\d \mathcal{V}$ as that
over
the system
surface and 
the normal coordinate
$\d \xi$,  $\d \mathcal{V}=J \sqrt{g}\d \eta \d \xi$,
where $\sqrt{g}$ is the Newtonian
volume element related to the metrics given in
Eq.~(\ref{metrictensor})
[see Eq.~(\ref{length}) for $\sqrt{g}$] and Eq.~(\ref{J})
for the 
Jacobian $J(r)$. For the spherical system, one has $\sqrt{g}=1$.
Then, we expand
the integration limits
of $\xi$ to 
$\pm \infty$ 
    because $(\partial \rho/\partial \xi)^2$
has  a sharp maximum at the ES ($\xi=0$).
Factor 2 appears because of Eq.~(\ref{eq0}) which leads to
the approximate (at leading order in small parameter $a/R$) equivalence
of the second order [$\eps^{}_{G}(\rho)$] correction to the
bulk energy density
$-b^{(G)}_{V}\overline{\rho}$, and the surface (gradient) parts of the
total energy $E$.
As we have square of density derivative over $\xi$ in Eq.~(\ref{sigma}),
one can use Eq.~(\ref{eq0}) 
for the density $\rho$ at the leading order.
 Equation (\ref{sigma}) can be transformed to the expression with
the dimensionless integral over $y$, that is  more convenient
for calculations,
\be\l{sigma1}
\sigma=-\frac{a \overline{\rho} K_{G}}{9} J(R)
\!\int_{0}^1 \!\d y (1+\beta y +\gamma/y)\frac{{\rm d} y}{{\rm d} x}
\approx\frac{a \overline{\rho} K_{G}}{9} J(R)
\!\int_{0}^1 \!\d y
\sqrt{\left(1 + \beta y +\gamma/y\right)\epsilon(y)},
\ee
where $J(R)$ is given by Eq.~(\ref{J}) at $r=R$.
In these derivations we transform the variable $\xi $ to $\rho$,
$\d \xi=(\d \xi/\d \rho) \d\rho$, as above, and use 
    Eqs.~(\ref{varepsilon}), (\ref{eq0}),
and (\ref{adef})-(\ref{consts}). 
For the approximation (\ref{epsilon}) to $\epsilon(y)$, the
integral in Eq.~(\ref{sigma1}) can be taken in terms of the
elementary functions. 
The simplest 
analytical
expression
for the tension coefficient $\sigma $ is obtained for $\beta=\gamma=0$
[case (i)].
 Using Eq.~(\ref{b0g0}) for $y=y(x)$, 
    one finally arrives at 
\be\l{sigma00}
\sigma=4a\frac{\overline{\rho} K_{G}}{135} J(R)~.
\ee
Similarly, the expression for $\sigma $ in the 
    Wilets case (ii) can be derived
 too,
$\sigma=a \overline{\rho}K_{G} J(R)\sqrt{\gamma}/36$.
Thus,
one obtains the explicitly analytical expressions for the
tension coefficient $\sigma$
as functions of constants of the energy density
$\mathcal{E}$,
Eq.~(\ref{enerden}), and Schwarzschild metric
    Jacobian at the ES, $J(R)$.
    For nuclear physics, the tension coefficient $\sigma$
    is related to the vdW-Skyrme interaction constants,
    in good agreement
with the van der Waals capillary theory\cite{RW82}.

 Equation (\ref{energytot})
is the two main (volume and surface) terms of the
usual leptodermic expression for the total energy of the dense liquid drop.
In nuclear physics, this expression was obtained
with including additional terms
of the symmetry energy [with volume (and Swiatecki linear term)
and surface (isovector) corrections] and 
curvature and Coulomb components (see, for instance,
Refs.~\citen{BM13,BM15} for the ESA).

Notice that due to 
  the integration over the normal-to-ES
  radial variable $\xi=r-R$,
  the smallness factor, proportional to $a/R$, appears for 
      the surface with
  respect to the volume contributions to the NS mass,
  Eq.~(\ref{MNSSfin}),
  and total energy $E$, Eq.~(\ref{enertot}).
  Thus, 
      these surface components, $M_{S}$ and $E_{S}$
      are of the order of a small parameter $a/R$. However, these surface
      contributions, $M_{S}$ and $E_{S}$,
      dramatically influences on the total NS characteristics
      because the vdW capillary surface pressure (including the
      gravitational forces) equilibrates the
      volume pressure acting from the interior of the  liquid-drop to
      be a leading reason of its equilibrium stability. We will
  study this stability condition in more details in the next section.

\begin{figure*}[!ht]
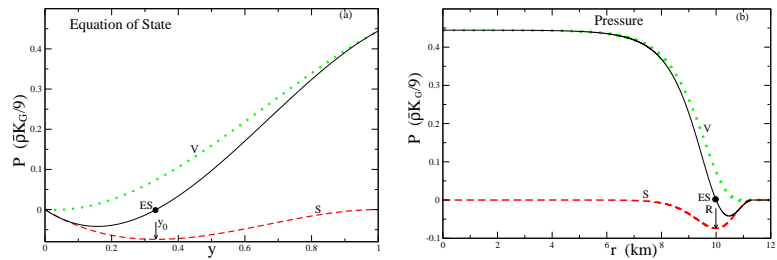

  \vskip1mm
 \centerline{\includegraphics[width=4.9cm]{Fig4a-PVStoty.eps}
  ~
    \includegraphics[width=4.9cm]{Fig4b-PVSr_a1R10.eps}}

  \vskip5mm\caption{{\small
      Pressure $P$,
       Eq.~(\ref{Pfin}), in units of 
      $ \overline{\rho} K_{G}/9$ 
       as function of
       the density variable $y=\rho/\overline{\rho}$ in (a)
       for the universal EoS,
      and of the radial
      coordinate $r$ in (b) 
      through the particle number 
          density $\rho=\overline{\rho}y((r-R)/a)$, where
      $y(x)$ is given, e.g., by Eq.~(\ref{b0g0})
          (solid lines). Dashed (red) and dotted (green) lines show the surface
          ($S$)
      and volume ($V$) components.
      The crust thickness $a=1$ km
      and the effective radius parameter $R=10$ km
      are the same as in Fig.~\ref{fig1}(b).
      The full dots and arrows present the ES for $y=y^{}_0=1/3$ (a)
      [see Eq.~(\ref{b0g0})]
      and $r=R$ (b).
}}
\label{fig4}
\end{figure*}

\vspace{1.0cm}
 \section{Equation of state}
 \l{eosP}

     The  equation of state (EoS), $P=P(\rho)$, can be determined\cite{LLv5} by
     using the energy density $\epsi(\rho)$, Eq.~(\ref{enerden}), 
\be\l{EoS}
P=\rho^2 \frac{\delta \mathcal{W}}{\delta \rho}~.
\ee
The energy per particle $\mathcal{W}$
     is a function of the particle number density $\rho$;
     see Eqs.~(\ref{satcond}) for $\mathcal{W}$, and (\ref{enerden})
     for the energy density $\epsi$.
    According to Eq.~(\ref{eq}) for the equilibrium of the
    dense liquid-drop system, one obtains
    the 
   pressure $P$ in the form:
\be\l{EoS1}
P=\mu \rho  -\mathcal{E}(\rho)=\mu \rho -\mathcal{A}(\rho)
- \mathcal{B}(\rho) (\nabla \rho)^2~,
\ee
where $\mu$ is the chemical potential, 
and $\mathcal{E}(\rho)$ is the energy density,
Eq.~(\ref{enerden}). The leading term of the
    pressure (\ref{EoS1})
can be presented as
\be\l{EoS2}
P\approx \mathcal{M}\rho-\varepsilon^{}_{G}(\rho)
- \mathcal{B}(\rho) \left(\frac{\partial \rho}{\partial \xi}\right)^2~,
\ee
where $\mathcal{M}=\mu+b^{(G)}_{V}$; see Eq.~(\ref{solvol}),
$\partial \rho/\partial \xi $ is given by Eq.~(\ref{eq0})
in terms of the $\varepsilon^{}_{G}(\rho)$, Eq.~(\ref{varepsilon}),
 $\mathcal{B} (\partial \rho/\partial \xi)^2=\varepsilon^{}_{G}(\rho)$.
For a leptodermic
system, $a/R \ll 1$,
one can
rewrite this equation as
\be\l{PvPs}
P\approx P_{V}+P_{S}~,
\ee
where $P_{V}$ is the volume part of the pressure
$P$ [see Eq.~(\ref{enerdenvol})]
\be\l{Pv}
P_{V} \equiv\mu \rho -\mathcal{A}(\rho)=\mathcal{M}\rho-
\varepsilon^{}_{G}(\rho)~.
\ee
The second component, $P_{\rm S}$, is the surface part,
\be\l{Ps}
P_{S}\equiv- \mathcal{B}(\rho) \left(\frac{\partial \rho}{\partial \xi}\right)^2
\approx -\varepsilon^{}_{G}(\rho)
\approx -\frac{\overline{\rho} K_{G}}{18} y(1-y)^2.
\ee
For the middle  we used Eq.~(\ref{eq0}) for
the particle number density $\rho$ in the leading order approximation.
Using Eqs.~(\ref{varepsilon}) and
(\ref{epsilon}) for the quadratic approximation, one arrives at the
last equation.
Notice that the surface component, $P_{S}$, is concentrated
near the effective surface, $r=R$, with the exponential
decrease inside and outside of the mass system; see Fig.~\ref{fig4}.
According to Eqs.~(\ref{Pv}) and (\ref{Ps}) for the quadratic approximation
to $\epsilon(y)/y$, Eq.~(\ref{epsilon}), one has a macroscopic
liquid-drop equilibrium
condition at the ES, $(P_{V}+P_{S})_{r=R}=0$. Thus, the chemical potential
surface component $\mathcal{M}$ at the liquid-drop
equilibrium equals
\be\l{Mcal}
\mathcal{M}=\frac{K_{G}}{9}(y^{}_0-1)^2,
\ee
where $y^{}_0$ is the value of the density $y(x)$ at the ES, $x=0$,
$y^{}_0=1/3$ for the vdW-Skyrme case (i), and
$y^{}_0=1/2$ for the Wilets solution (ii). With Eq.~(\ref{Mcal}), one finally
obtains from Eqs.~(\ref{PvPs}), (\ref{Pv}), and (\ref{Ps}),
\be\l{Pfin}
P\approx \frac{K_G}{9}~y\left[(y^{}_0-1)^2-(y-1)^2\right]=P_{\rm V}+P_{\rm S}~,
\ee
where
\bea\l{Pvfin}
&P_{V}=\frac{K_{G}}{9}~y\left[(y^{}_0-1)^2-\frac{1}{2} (y-1)^2)\right]~,
\\
&P_{S}=-\frac{K_{G}}{18}~y (y-1)^2~.
\l{Psfin}
\eea

 Figure~\ref{fig4} shows the 
 pressure $P$ as function of the dimensionless density $y$, (a),
 and radial variable $r$, (b), through the density, $y((r-R)/a)$,
at leading order over
the parameter $a/R$. See Eq.~(\ref{Pfin}) for the
full pressure,
Eq.~(\ref{Pvfin}) for the volume ($P_{V}$), and Eq.~(\ref{Psfin})
for the surface ($P_{\rm S}$)
components of the pressure. As seen from Fig.~\ref{fig4}(a),
all pressures, $P$, $P_{V}$ and $P_{S}$, take zero value at $y=0$.
The surface pressure, $P_{S}$, is zero
also asymptotically
in the limit $y \rightarrow 1$, in contrast to the volume part, $P_{V}$
(the total pressure $P$). The full circle shows the
ES at $y=y^{}_0=1/3$ for
the (i) case. As expected, the volume pressure  $P_{V}$ is monotonically
increasing function of the density $y$. In Fig.~\ref{fig4}(b), the
pressure
$P(\rho(r))$ (solid) has a typical leptodermic behavior with a relatively
short thickness of the order of $a$,  where the pressure decreases sharply
from almost the constant, which is the asymptote
inside of the system at
$r \simg R-a$,  to zero. 
Dashed red and dotted green lines show
the surface ($P_{S}$) and volume ($P_{V}$) components, respectively.
Sum of these components is the total pressure $P$ with zero at the ES;
see
the full point (ES) and arrow which present the effective surface and
radius $R$, respectively.
As seen also
from Fig.~\ref{fig4}(b), 
we obtain analytically
a sharp minimum of the surface pressure, $P_{\rm S}(r)$,
near the effective surface
($r \approx R$)
with 
 its exponential decrease 
 inside and outside of the ES to zero. 
 In the
  vdW-Skyrme
case (i), the pressure $P(r)$ is exponentially sharp decreasing function of
$r$ outside of the system.
Notice that, in contrast to this, the Wilets
solution in the symmetric case
(ii) for the particle density $\rho$, Eq.~(\ref{wilets}), leads
to a large tail outside of the system 
    as expected from Fig.\ \ref{fig1}(b).  
For the
parameters 
$R=10$ km, and $a=1$ km (see Figs.~\ref{fig4} and \ref{fig1}(b)),
one finds that the pressure $P(r)$ [Eq.~(\ref{Pfin})] disappears  in
the vdW-Skyrme case (i) at 
$r\simg R+a x_0$ ($x_0\approx 1.32$), that corresponds  to about
$r \simg 11.3 $ km, similarly as for the density $\rho(r)$,
which is a very close distance 
from the ES. 
This  distance is
decreased much for the pressure $P(r)$
in the case of the significantly small edge thickness $a=0.1$ km to 10.7 km,
that is somehow larger
than for the
particle number density.

\section{Tolman-Oppenheimer-Volkoff approach}
\l{tov}

The basic relations determining the mass-radius
relation are the traditional TOV
differential equations\cite{RT39,OV39,RT87,SBL23}:
\be\l{toveq}
\frac{{\rm d} P}{{\rm d} r}= 
-\frac{G(\epsi + P )(mc^2 + 4\pi r^3P )}{rc^2(rc^2 - 2G m)} ,\quad
\frac{{\rm d} m}{{\rm d} r}
= \frac{4\pi r^2\epsi}{c^2}~,
\ee
where $m(r)$ is the gravitational mass interior to the radius $r$,
$\epsi$ is the energy density, 
Eq.~(\ref{enerden}) in our approach.
     Boundary conditions for these equations are
     \be\l{boundcondmp}
     m(r = 0) = 0, \quad (\d P/\d r)_{r=0} = 0~~.
\ee
     The integrations are terminated when
     $P = 0$, which defines the surface $r = R$. A specified value
     of the
     central pressure $P_0=P(r = 0)$ determines the total mass
     $M = m(r = R)$ in the TOV approach.
     Note that the only EoS relation needed
     is the pressure 
     density
     relation $P=P(\rho)$.

     Taking approximately into account the step-function density,
     $\rho=\overline{\rho}$ for $r < R$ inside of the system, and $\rho=0$ for
     $r > R$ outside of it,
          one can solve TOV equations (\ref{toveq})
     analytically. Our leading-order solution,
     Eq.~(\ref{eq0yx}), and examples,
Eqs.~(\ref{b0g0}) and
(\ref{wilets}) [see Fig.~\ref{fig1}(b)], 
having a sharp transition from the saturation to zero
value, largely agree with above mentioned
approximation at small leptodermic parameter
$a/R \ll 1$, the better the smaller this parameter.
The step-like density
can be considered as the zero order in the leptodermic expansion in the limit
$a/R \rightarrow 0$. 
Taking into account the first boundary condition
     in Eq.~(\ref{boundcondmp}), the second equation in
     Eq.~(\ref{toveq}) can be integrated
     analytically, 
     \be\l{mres}
     m(r)=\frac{4\pi\epsi_0}{3 c^2}r^3,
     \ee
     where
     \be\l{Eps0}
     \epsi_0=\epsi(\rho=\overline{\rho}).
     \ee
     This value should be some constant, independent of $r$ for the
     assumed coordinate dependence of the density $\rho(r)$ 
         inside of
     the particle system. In particular,
     it can be taken approximately as shown in 
Eqs.~(\ref{enerden}), (\ref{vol}), and (\ref{varepsilon}).
Therefore,
the total mass $M$ in this approach is given by
     \be\l{Mtot}
     M=
     m(R)=\frac{4\pi\epsi_0}{3 c^2}R^3,
     \ee
     where $R$ is the effective NS radius ($r_1$ in the notations of
     Tolman's book, Ref.~\citen{RT87}).  Notice that the NS mass $M$,
     Eq.~(\ref{MNSSfin}), derived in Section \ref{NSmass}, differs
     essentially from this expression, Eq.~(\ref{Mtot}), by the surface
     component $M_{\rm S}$; see Eq.~(\ref{MNSSfin}).
     Another essential
     difference comes from the Schwarzschild metric Jacobian $J$,
     Eq.~(\ref{J}), with
     the influence on the surface and volume parts of the NS mass.

     Substituting Eq.~(\ref{mres}) into the first TOV Eq.~(\ref{toveq}),
     one can see that the variables $P$ and $r$ are separated. Therefore,
     we
     analytically find  the solution
     for $r \leq R \leq  R_{\rm SM}$ and
         $(P+\epsi_0/3)/(P+\epsi_0)>0$: 
     \be\l{solTOVplus}
         P = \frac{\epsi_0}{3}
    \frac{3\zeta \sqrt{1-r^2/R_{\rm SM}^2}-1}{1-
     \zeta\sqrt{1-r^2/R_{\rm SM}^2}},
     \ee
     where
     \be\l{zeta}
     \zeta=\Big |\frac{P_0 + \epsi_0/3}{P_0 + \epsi_0}\Big |~,
     \ee
     and $P_0\equiv P(\overline{\rho})=P(r=0)$. 
     In Eq.~(\ref{solTOVplus}), $R_{\rm SM}$ is the radius parameter of the 
     Schwarzschild metric in the form presented in
     Eq.~(\ref{Schwarz});
         see
    Eq.~(\ref{RTOV}) for $R_{\rm SM}$, and Ref.~\citen{RT87}.
         This metric
     is non-singular  at $r=0$. It was
    derived from
    the original form, valid outside of the NS matter system,
    Refs.~\citen{RT87,LLv2}, and singular at $r=0$,
    by using a transformation of coordinates
    and GRT invariance; see, e.g., Ref.~\citen{RT87}.  

\begin{figure*}[!ht]
  \vskip1mm
  \centerline{\includegraphics[width=5.9cm]{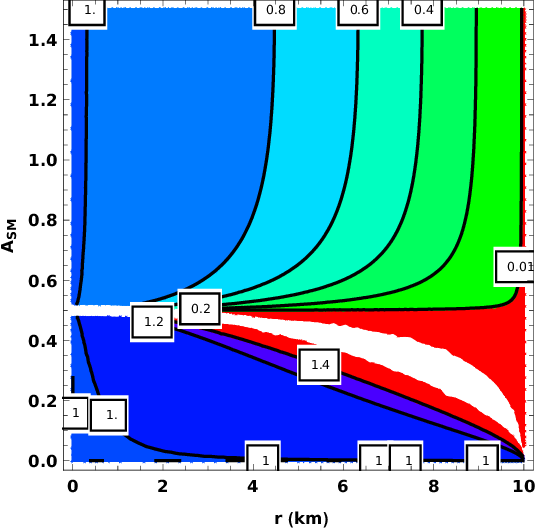}}
  
\vspace{-0.4cm}
  \caption{{\small
      Contour plots for the
       pressure $P(r)$, Eq.~(\ref{Tolsol}),
      in units of 
     the central value, $P(r=0)$, 
      as function of the radial
      coordinate $r$ and the dimensionless
      parameter $A_{\rm SM}$ of the 
          Schwarzschild metric,
      Eq.~(\ref{Schwarz}). The numbers in squares show the values of
      this pressure.  White color presents regions where we have
          indetermination, infinity by infinity, with the finite limit 1
          at $r \rightarrow 0$. Red color shows negative values of the
           ratio $P(r)/P(r=0)$ 
          [a positive pressure,
          $P(r)$]. 
          The value
          ``0.01'' displays approximately the zero value in
          horizontal and
      vertical lines on right of plots.
      The effective NS radius $R=10$ km is the same as in Figs.~ \ref{fig4} and
      \ref{fig1}(b).
 }}
\label{fig5}
\end{figure*}
\begin{figure*}[!ht]
  \vskip1mm
   \centerline{\includegraphics[width=5.9cm]{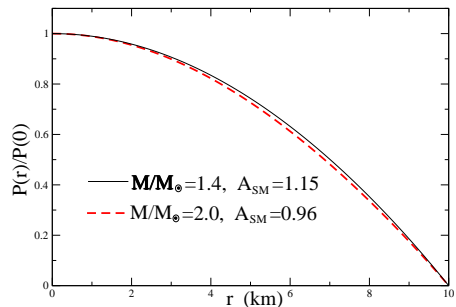}}
  
\vspace{-0.4cm}
  \caption{{\small
    The pressure $P(r)$, 
      Eq.~(\ref{Tolsol}),
      in units of 
     the central value, $P(r=0)$ 
      as function of the radial
      coordinate $r$ for the Schwarzschild
      metric in the form Eq.~(\ref{Schwarz}) at two 
          values of the parameter $A_{\rm SM}$, $A_{\rm SM}=1.15$
      (solid line)
          and $A_{\rm SM} \approx 0.96$ (dashed line) which correspond
          approximately to
          the two limit 
          values for the NS masses, $M=1.4 M_\odot$
          and $M=2.0 M_\odot$, which are related to the recent
          experimental data\cite{GR21,CC20,TR21}
          ($R_{\rm SM} \approx 13.0$ km and
      $15.6 $ km, respectively ; see text).
        The effective NS radius $R=10$ km is the same as in Fig.~ \ref{fig4}. 
}}
\label{fig6}
\end{figure*}
    Notice that the solution (\ref{solTOVplus})
     coincides, up to a constant related to units, with
     that presented in Tolman's book (Ref.~\citen{RT87})
         at $\zeta=B_{\rm SM}/A_{\rm SM}=1/(2A_{\rm SM})$, 
     \be\l{Tolsol}
     P=\frac{\epsi_0}{3}~
     \frac{3 B_{\rm SM}\sqrt{1-r^2/R^2_{\rm SM}} - A_{\rm SM}}{
       A_{\rm SM}-B_{\rm SM}\sqrt{1-r^2/R^2_{\rm SM}}}~,
     \ee
    where
     \be\l{BAzeta}
     B_{\rm SM}=\frac{1}{2}, \quad A_{\rm SM}=\frac{1}{2\zeta}~.
     \ee
     We have to choose the sign plus on right of
         last equation     
     for the positive expression of
         $(P+\mathcal{\epsi}_0/3)/(P+\mathcal{\epsi}_0)$, 
     because the  minus, which is related to the opposite sign of the
     same expression,
     is not valid for the Schwarzschild metric, $A_{\rm SM}>0$.     
    The cosmological constant is
     assumed to be zero, 
    $\epsi_0 \propto R^{-2}_{\rm SM}$, Eq.~(\ref{RTOV}),
     ($G=c=1$ in Tolman's book).
        The constants, $A_{\rm SM}$, $B_{\rm SM}$, and $R_{\rm SM}$ are
     parameters of the 
     transformed interior Schwarzschild 
         metric, Eq.~ (\ref{Schwarz}), for $A_{\rm SM}>0$,
     \be\l{Tolconsschwm}
      A_{\rm SM}=\frac32\sqrt{1-\frac{R^2}{R_{\rm SM}^2}},\quad
      B_{\rm SM}=\frac12~.
      \ee
             Therefore, one obtains the relation of the parameters of the
      Schwarzschild metric, Eq.~ (\ref{Schwarz}), to the initial condition
      $P=P_0$ at $r=0$ through Eq.~(\ref{zeta}).          
          Notice that according to Eq.~(\ref{Tolsol}) for the pressure
          $P(r)$, one has $P=0$ at the boundary $r=R$
      because of
      Eq.~(\ref{Tolconsschwm}) for the constants $A_{\rm SM}$ and $B_{\rm SM}$.
        Finally, one has 
         the full agreement of
      our result for the pressure,  
          Eq.~(\ref{solTOVplus}),
       with that of
     Ref.~\citen{RT87}.

\vspace{1.0cm}
 \section{Discussions of the results}
 \l{discres}

    Figure~\ref{fig5} shows the contour plots for the pressure
     $P(r,A_{\rm SM})$
      [in units of 
     $P(r=0,A_{\rm SM}$)],
     as function of $r$ (km) and parameter $A_{\rm SM}$ of the
     Schwarzschild metric (\ref{Schwarz}); see 
     Eq.~(\ref{Tolconsschwm}). 
         It is more convenient to use  the finite
     dimensionless  values of $A_{\rm SM}$, $ 0 \leq A_{\rm SM}\leq 1.5$,
     instead of dimensional $R_{\rm SM}$ in the infinite range
     for the fixed effective  NS surface
     radius
     $R=10 $ km, $0\leq r \leq R $. For a rough estimate
     $\epsi_0 \sim c^2\rho_c$, where the central mass density
     $\rho_c \approx M/(4\pi R^3/3)$ 
     ($M\approx M_{V}$), and $M/M_{\odot} \approx 1.4-2.0$
     in units of the Solar mass $M_{\odot}$, 
     one finds significantly larger parameter $R_{\rm SM}$ than
     the effective radius $R$, $R_{\rm SM} \approx 15.6-13.0$ km in
     this case\footnote{
     Notice that the Schwarzschild
     parameter $R_{\rm SM}$ differs essentially
        from the gravitational radius $r_{\rm g}$, introduced by
        Schwarzschild,
        $r_{\rm g}=2GM/c^2 \leq R$; see Ref.~\citen{LLv2}.
        Indeed,  $R_{\rm SM}$ is assumed to be 
        larger
          than 
         the effective NS radius $R$.}, respectively.
     With these
     evaluations, one has respectively
     the constants $A_{\rm SM} \approx 1.15-0.96$
     for $M 
     \approx (1.4-2.0) M_{\odot}$. Equation (\ref{BAzeta})
         presents the definite relation between this $A_{\rm SM}$
         and the dimensionless 
         variable $\zeta$, Eq.~(\ref{zeta}), which can be also considered
         equivalently instead of 
         the two dimensional values of the initial pressure 
         $P_0$ and energy density $\epsi_0$. 
     For the gravitational radius, one obtains
     $r_{\rm g}=4.1-5.9$ km,
        that is smaller than the effective radius $R=10$ km. 
        The maximum value of $A_{\rm SM}$,
        $A_{\rm SM} = 1.5$, corresponds to
        zero pressure $P(r)$.

                 Figure~\ref{fig6} shows the 
            two cuts of the contour plots
        (Fig.~\ref{fig5}) at the value  $A_{\rm SM}= 1.15$
        ($R_{\rm SM}\approx 15.6$ km,
        solid black line),
        and the value
        $A_{\rm SM}=0.96$ ($R_{\rm SM} \approx 13.0$ km,  dashed red line). 
        The exemplary physical region, $ M/M_\odot 
        \approx 1.4-2.0$,
               which are consistent
            with the recent experimental
            data shown in Fig.~\ref{fig2} for NS masses
            (see Refs.~\citen{GR21,CC20,TR21}, also
            Refs.~\citen{BZ14,HPY07,ST04,OL20}),
            is related to
            a small range
            in Fig.~\ref{fig6}. These curves have the same starting
            ordinate 1 and final one 0~.

            For more realistic values of $R_{\rm SM}$, for instance,
     taking into account the gravitational defect of mass\cite{LLv2} in
     the evaluations of the central energy density $\epsi_0$, one may have
     different values of $A_{\rm SM}$ presented in Fig.~\ref{fig5}.
     Thus, as seen 
         from solid and dashed lines of Fig.~\ref{fig6} and those assumed in
     the derivations of the TOV equations\cite{OV39,RT87}, the
       pressure
         $P$ 
         turns, indeed, to zero
         at the boundary
         of the NS system, $r=R$.

\begin{figure*}[!ht]
  \vskip1mm
 \centerline{\includegraphics[width=7.9cm]{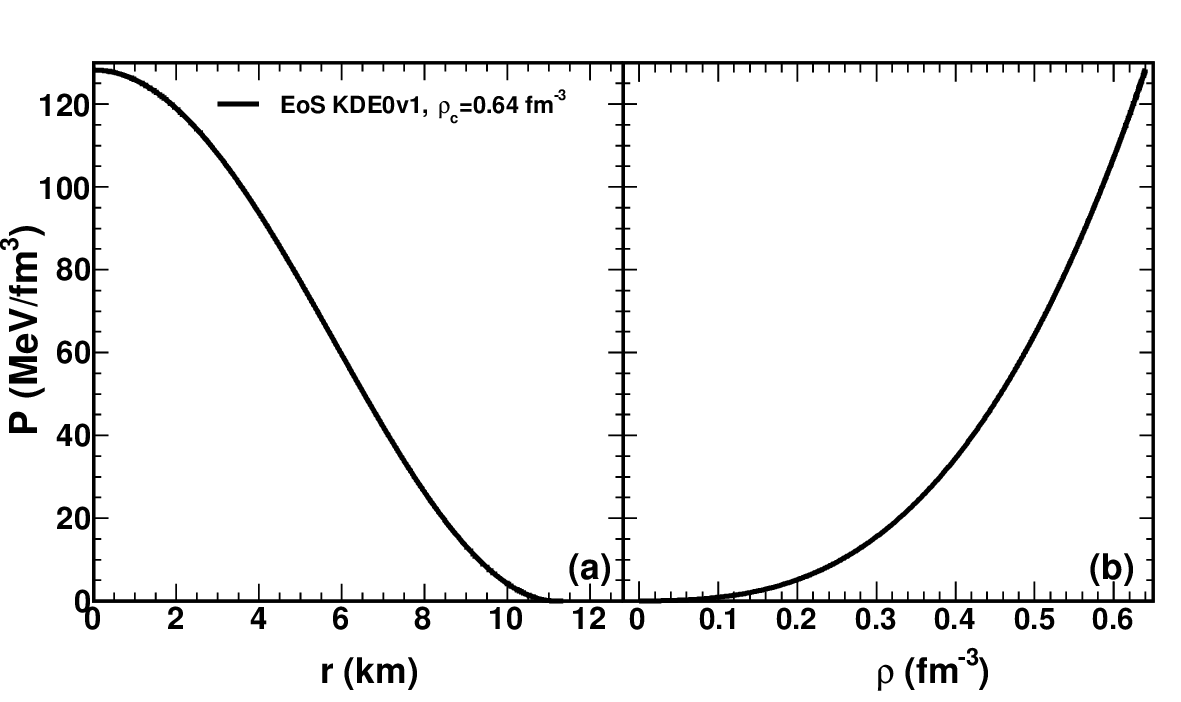}}

  \vspace{-0.5cm}
  \caption{{\small
      Left (a): The pressure versus r of the NS with the
      Skyrme interaction KDE0v1 by solving TOV equations numerically.
      Right (b): The pressure versus the corresponding density of the NS.
      The central density $\rho_c=0.64$ fm$^{-3}$ and the mass of the NS
      is 1.62 M$_\odot$.
      }}
\label{fig7}
\end{figure*}

\begin{figure*}[!ht]
  \vskip1mm
  \centerline{\includegraphics[width=4.9cm]{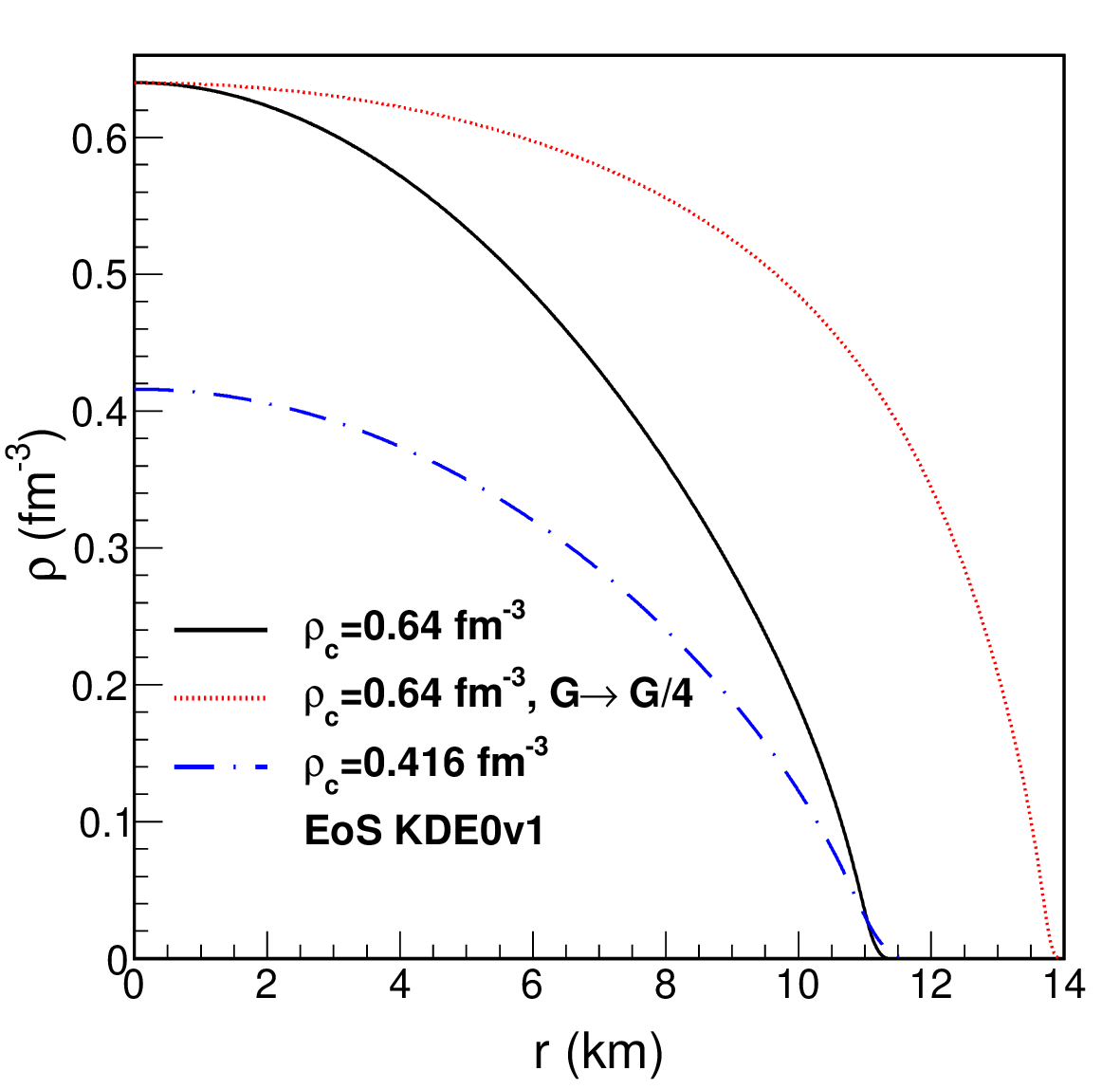}
  ~
   \includegraphics[width=4.9cm]{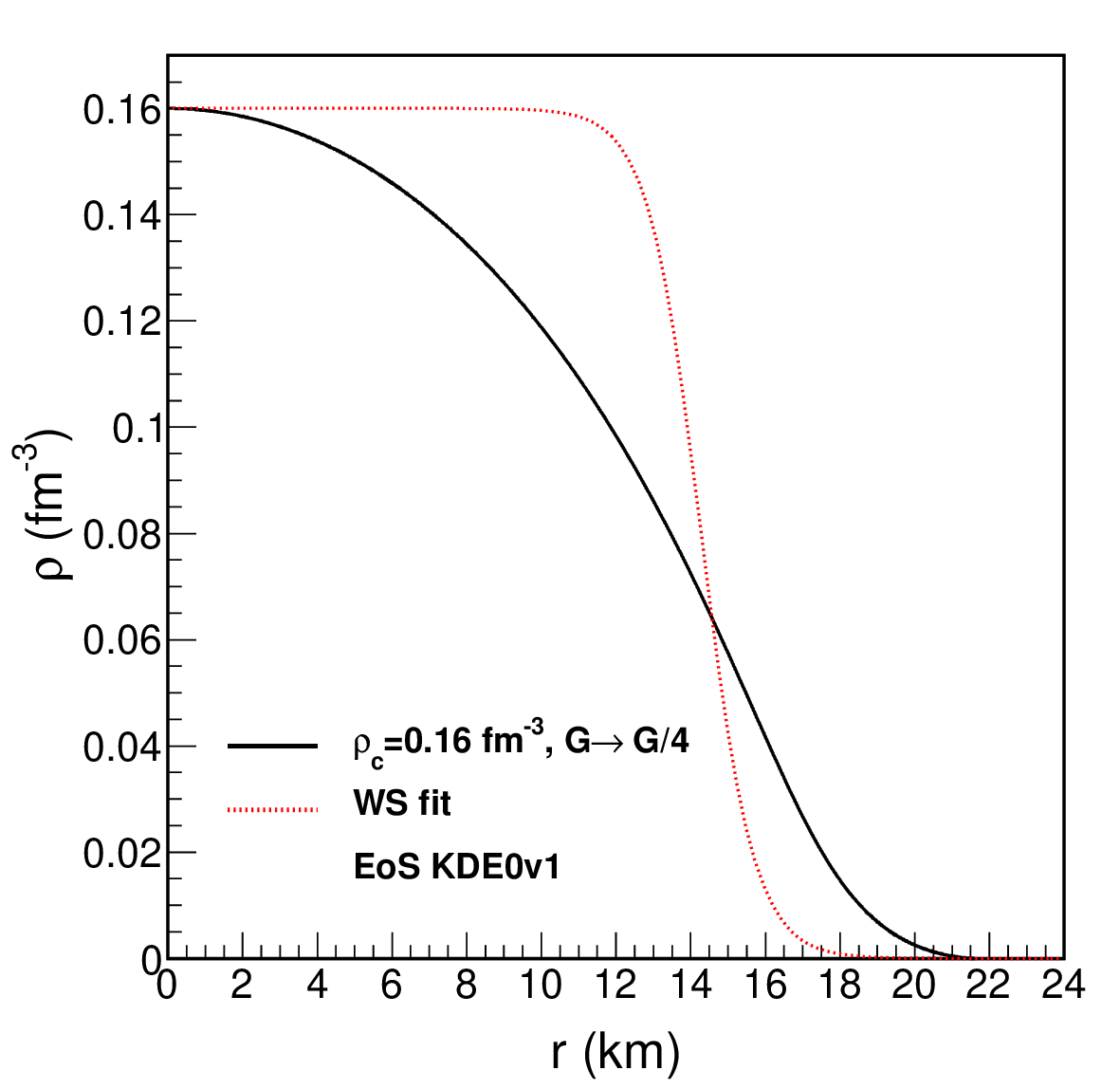}}

  \vspace{-0.5cm}
  \caption{{\small
     {\it Left:} The density profiles of the NS with the Skyrme interaction KDE0v1
      by solving TOV equations numerically. The black solid line refers to the
      central density $\rho_c=0.64$ fm$^{-3}$ and the mass of the NS is
      1.62 M$_\odot$. The red dotted line refers to the central
      density $\rho_c=0.64$ fm$^{-3}$ with changing
      the gravitational
      constant G to G/4 and the mass of the NS is 4.0
      M$_\odot$. The blue dashed-dotted line refers to another
      central density $\rho_c=0.416$ fm$^{-3}$ and the mass of the NS
      is 1.02 M$_\odot$.
      {\it Right:} The density profiles of the NS with the
      Skyrme interaction KDE0v1
      by solving TOV equations numerically. The black solid line refers to
      the central density $\rho_c=0.16$ fm$^{-3}$ with changing the
      gravitational constant G to G/4 and
      keeping the mass of the NS to be 1.62 M$_\odot$. The red dotted line
      is the WS fit, Eq.~(\ref{WSfit}); see the text. 
      }}
\label{fig8}
\end{figure*}

Figure~\ref{fig7} shows the results obtained
numerically by solving the TOV
     equations (\ref{toveq}) for the pressure $P=P(r)$ (a) with the EoS,
     $P=P(\rho)$ (b), related to the Skyrme force
     KDE0v1 \cite{AS05}; 
     see Refs.~\citen{AS05,BZ14,ZB11}. 
     For convenience, the radial coordinate $r$ in (a)
     is presented in km units while the density variable $\rho$
     is done in nuclear units fm$^{-3}$. In both panels (a) and (b), the same
     pressure $P$ is displayed in nuclear units MeV/fm$^3$.
     The pressure $P(r)$ decreases with increasing $r$ while
     the EoS function of $\rho$, $P=P(\rho)$, has an opposite behavior.
     These numerical semi-microscopic results are qualitatively
      in agreement with our macroscopic
     volume components
    $P_V$ shown in Fig.\ \ref{fig4} inside of the NS.

     The particle number density $\rho$ is shown  in Fig.\ \ref{fig8}(a) as
     function of the radial coordinate $r$ (km) for different parameters.
      The masses $M/M_{\odot}=1.62$ and $1.02$, larger than the Solar mass
     $M_{\odot}$, correspond to different values of the central density
     $\rho_c=0.64$ fm$^{-3}$ (solid black) and
     $0.416$ fm$^{-3}$ (blue
     dashed-dotted curve), significantly larger than the nuclear matter
     value $0.16$ fm$^{-3}$. All of them are related to the EoS
     KDE0v1 Skyrme force\cite{AS05}; see also Refs.~\citen{BZ14,ZB11}.
      The gravitational force effects are tested
      formally for decreasing the gravitational
      constant $G$ by factor 4
     in red dotted line in Fig.\ \ref{fig8}(a).
     The Woods-Saxon leptodermic
     fit is shown by the red dotted line
     in Fig.\ \ref{fig8}(b).
     The last situation is qualitatively close to our analytical results
     for the volume (V) contributions presented in
     Figs.~\ref{fig4}-\ref{fig6}.
     The main difference in Fig.~\ref{fig4}(b), as compared to the numerical
     results in
     Figs.~\ref{fig7},\ref{fig8}, is the surface contribution (S) into EoS,
     $P=P(y)$ ($y=\rho/\overline{\rho}$) [see
     Fig.~\ref{fig4} (a)] and
     the pressure leptodermic correction $P_{\rm S}(r)$,
     Eq.~(\ref{Psfin}) [Fig.~\ref{fig4}(b)].
         The shape of the total
     pressure behavior, $P=P(r)$, is more
     leptodermic in Fig.~\ref{fig4}(b) than in that of
     Fig.~\ref{fig7} (left). These results are in good agreement with
     the density shapes for
     the analytical results in Fig.~\ref{fig1}(b) and numerical
     calculations in
     Fig.~\ref{fig8} (right) as displayed by the red curve for the 
         inversed Woods-Saxon
     (WS) fit.
     The WS fit formula is given by
\begin{equation}
   \rho(r)=
  \frac{\rho^{}_{\rm WS}}{1+
    \exp\left[\left(r-R_{\rm WS}\right)/a^{}_{\rm WS}\right]}~,
  \l{WSfit}
\end{equation}
where $\rho^{}_{\rm WS}$=0.16 fm$^{-3}$, $R_{\rm WS}=14.2786$ km, and
$a^{}_{\rm WS}=0.7074$ km.
    The results in Figs.~\ref{fig7}(a) and \ref{fig8} (right, red dotted line)
    for the volume contributions of the pressure $P(r)$,
    obtained analytically from the TOV equations (\ref{toveq}),
    are naturally
     similar to
     those shown
     in Fig.~\ref{fig6}, and Fig.~\ref{fig4} (b) for the volume contribution.
     The surface contributions to the standard
     TOV equations (\ref{toveq})
     and the corresponding solutions will be studied in the forthcoming
     work where we are going to take into account explicitly
     the ES corrections.

     \section{Conclusions}
\l{concl}
     
     The effective surface approximation based on the leptodermic expansion
     over a small parameter - a ratio of the crust thickness $a$
     to the effective radius $R$ of the system - is extended for
     the description
      of neutron star properties.
       Neutron star was considered  as 
 a dense liquid drop at
 the equilibrium.   The gravitational potential
  $\Phi(\rho)$ was taken into account 
    in the simplest form as expansion over powers of
 differences $\rho-\overline{\rho} $, where
        $\overline{\rho}$ is the saturation density,
 up to second order in terms of the separation particle energy and
 incompressibility. For a strong gravitation, one has its significant
 contribution to the  separation particle energy $b^{(G)}_{V}$ and
 incompressibility modulus $K_{G}$,
 within the Schwarzschild
 metric solution to the General Relativity Theory equations.
 Taking into account the gradient terms of the energy
 density in a rather
 general form we
 analytically obtained the leading (over a small parameter $a/R \ll 1$)
 particle
 number density $\rho$
 as function of the normal-to-ES coordinate
 $\xi$ of the orthogonal local nonlinear-coordinate system
 $\xi, \eta, \varphi$
 through the effective surface. 
     This result is in a good agreement
     with the van der Waals phenomenological capillary theory. With the help
 of the local coordinate system $\xi,\eta$, one finds
  the equation of state (EoS) for the pressure $P=P(\rho)$, and the NS 
  surface mass and energy corrections 
  for a spherical 
  NS in the leading
      ES approach. 
          Our results for the dependence of the NS mass $M$ on
          the NS radius $R$, including the gravitational effects
          through the Schwarzschild metric,  are in a
          reasonable agreement with their
          recent experimental data for
      several neutron stars.
  Adding a
  first order correction over $a/R$, one obtains also the NS energy
 as a sum of the volume and surface terms
  with the analytical expression for
  the surface tension coefficient, also in agreement with the
  van der Waals theory. We obtained the effective
      surface corrections to the
      pressure $P$ of the EoS, which are significant near the ES for
      macroscopic condition
      of the NS stability. In line
  of Tolman derivations\cite{RT87},
  for the step-like particle number density, the
  TOV equations were solved analytically 
      in terms of the initial values of
  the pressure and energy density at  zero radial coordinate.
       The volume contributions to the pressure
       in our macroscopic
       analytical calculations are in good agreement with the
 semi-microscopic numerical results obtained from the TOV equations with the EoS based on the
  Skyrme forces, in the case of leptodermic fit of the particle number
  density $\rho$ to 
      its behavior for a Woods-Saxon potential.
      From comparison of the analytical
  and numerical results for the pressure and density, one can conclude
  importance of the surface corrections near the 
      NS effective surface.

  As perspectives, one can generalize  our analytical approach to 
  take into account the
  gradient terms in the TOV equations, also
   to 
  many-component and rotating systems.
  It is especially interesting to extend
  this approach to take into account  the symmetry energy
  of the isotopically asymmetric finite systems.
  As this approach was formulated in the local nonlinear system of coordinates
  $\xi, \eta, \varphi$ for deformed and superdeformed shapes of the
  effective surface,
  one can apply our method to the NS rotating
  pulsars at
  large angular momenta.

\section*{Acknowledgements}

The authors greatly acknowledge  C.A. Chin, V.Z.\ Goldberg, 
    S.N. Fedotkin,
J.\ Holt, C.M.\ Ko, E.I.\ Koshchiy, J.B.\ Natowitz, A.I.\ Sanzhur,
G.V.\ Rogachev 
for many creative and useful discussions.

\appendix

\section{Local coordinates near the effective surface}
\l{appA}
The axially-symmetric shapes of the ES
in cylindrical coordinates
 are assumed to be 
determined in terms of a certain profile function $y=Y(z)$ in 
the plane of axis of symmetry by rotation around $z$ axis. 
The $\xi$ is defined in the text as 
the coordinate  perpendicular to the ES.
The other coordinate $\eta$ can be 
chosen for example, as $z$-coordinate of the point at the surface where the 
perpendicular to the ES
from the given point ${\bf r}$ crosses it. Thus, let us define
new coordinates $\xi$,$\eta$, related
to the cylindrical coordinates, as
\be\l{transform}
y=y(\xi, \eta)=Y(\eta) +\frac{\xi}{{\cal L}},~~~
z= z(\xi,\eta)=
\eta - \frac{\xi}{{\cal L}} \frac{\partial Y(\eta)}{\partial \eta},
\ee
where
\be\l{calL}
{\cal L}= \left[1+ \left( \frac{\partial Y(\eta)}{\partial \eta}\right)^2
  \right]^{1/2}.
\ee
The length and volume elements are given by
\be\l{length}
\d {\it l} = \left(g_\xi \d \xi^2 +g_\eta \d\eta^2 + g_\varphi \d \varphi^2
  \right)^{1/2},~~
  \d {\bf r}= \sqrt{g}\d \xi \d \eta \d \varphi,~~~
  \sqrt{g}=
\left(R_1 +\xi\right)\left(1+ \xi/R_2\right)
\ee
with the diagonal metric tensor,
\be\l{metrictensor}
g_\xi=1,\qquad g_\eta=(1 + \frac{\xi}{R_2})^2 {\cal L}^2,\quad
g_\varphi=\left(1 + \frac{\xi}{{\cal L}}\right)^2 Y^2(\eta),
\ee
$R_1$ and $R_2$ are two local curvature radii of the ES,
\be\l{radii}
R_1={\cal L} Y(\eta), \qquad\qquad 
R_2=-\frac{{\cal L}^3}{\partial^2 Y/\partial \eta^2}~.
\ee
According to the right of Eq.~(\ref{length}),
the mean curvature writes
\be\l{curvature}
H=\frac{1}{2}\left(\frac{1}{R_1}+\frac{1}{R_2}\right)=
\frac{1}{2}\left(\frac{\partial }{\partial \xi} \ln \sqrt{g}\right).
\ee
For the Laplace operator in new coordinates $\xi$ and $\eta$, one finds
\be\l{laplacian}
\Delta_{\xi\eta} =\frac{\partial^2}{\partial \xi^2} +
\frac{\partial }{\partial \xi}\left(\ln \sqrt{g}\right)\; 
\frac{\partial }{\partial \xi} + \frac{1}{\sqrt{g}}
\frac{\partial }{\partial \eta}
\left(\frac{\sqrt{g}}{g_\eta}\frac{\partial}{\partial \eta}\right).
\ee

ORCID
\\
A. G.\ Magner https://orcid.org/0000-0003-1694-640X
\\
S. P.\ Maydanyuk https://orcid.org/0000-0001-7798-1271
\\
A.\ Bonasera https://orcid.org/0000-0001-7147-4535
\\
H.\ Zheng https://orcid.org/0000-0001-5509-4970
\\
A. I.\ Levon https://orcid.org/0000-0002-9880-1927
\\
T. M.\ Depastas https://orcid.org/0000-0001-7147-4535
\\
U. V.\ Grygoriev https://orcid.org/0000-0002-2684-2586

\end{document}